\documentclass[runningheads]{llncs}
\usepackage{amsmath,amssymb}
\usepackage{graphicx,pgfplots}
\usepackage[font=small]{subfig,caption}
\usepackage{algorithm2e}
\usepackage{todonotes,lineno,paralist,soul}
\usepackage{rotating,hyperref,multicol}
\usepackage{booktabs,tabularx}
\graphicspath{{figures/}}

\tikzstyle{densely dashed}=[dash pattern=on 0.75pt off 0.5pt]
\pgfplotsset{ytick style={draw=none}}

\pgfkeys{/pgf/number format/.cd,1000 sep={}}
\pgfplotscreateplotcyclelist{mylist}{%
{black,thick,smooth},
{black!50!white,thick,smooth},
{black,thick,densely dashed,smooth},
{black!50!white,thick,densely dashed,smooth},
{black,double,smooth}
}
\pgfplotsset{
        table/search path={data},
    }
\pgfplotsset{
  /pgfplots/xlabel near ticks/.style={
     /pgfplots/every axis x label/.style={
        at={(ticklabel cs:0.5)},anchor=near ticklabel
     }
  },
  /pgfplots/ylabel near ticks/.style={
     /pgfplots/every axis y label/.style={
        at={(ticklabel cs:0.5)},rotate=90,anchor=near ticklabel}
     }
}
\newcommand{\myparagraph}[1]{\smallskip\noindent\textbf{#1}.}

\newcommand{\ver}{arxiv}
\newcommand{\arxapp}[2]{\ifthenelse{\equal{\ver}{conf}}{#2}{#1}}

\author{Michael~A.~Bekos, Henry~F\"orster, Christian~Geckeler, Lukas Holl\"ander, Michael~Kaufmann, Amad\"aus~M.~Spallek, Jan~Splett}

\authorrunning{M.~A.~Bekos et al.}
\title{A Heuristic Approach towards Drawings of Graphs with High Crossing Resolution}
\titlerunning{Drawings of Graphs with High Crossing Resolution}

\institute{
Institut f\"ur Informatik, Universit\"at T\"ubingen, T\"ubingen, Germany\\
\texttt{\{bekos,foersth,geckeler,mk\}@informatik.uni-tuebingen.de}\\
\texttt{\{jan-lukas.hollaender,amadaeus.spallek,jan.splett\} @student.uni-tuebingen.de}
}

\begin{document}
\maketitle

\begin{abstract}
The \emph{crossing resolution} of a non-planar drawing of a graph is the value of the minimum angle formed by any pair of crossing edges. Recent experiments have shown that the larger the crossing resolution is, the easier it is to read and interpret a drawing of a graph. However, maximizing the crossing resolution turns out to be an NP-hard problem in general and only heuristic algorithms are known that are mainly based on appropriately adjusting force-directed algorithms.

In this paper, we propose a new heuristic algorithm for the crossing resolution maximization problem and we experimentally compare it against the known approaches from the literature. Our experimental evaluation indicates that the new heuristic produces drawings with better crossing resolution, but this comes at the cost of slightly higher aspect ratio, especially when the input graph is large.
\end{abstract}

\section{Introduction}
\label{sec:introduction}

In Graph Drawing, there exists a rich literature and a wide range of~techniques for drawing planar graphs; see, e.g.,~\cite{DBLP:journals/combinatorica/FraysseixPP90,DBLP:conf/gd/GutwengerM98,DBLP:journals/algorithmica/Kant96}. However, drawing a non-planar graph, and in particular when it does not have some special structure (e.g., degree restriction), is a difficult and challenging task, mainly due to the~edge crossings that negatively affect the drawing's quality~\cite{DBLP:journals/iwc/Purchase00}. As a result, the established techniques are significantly fewer (e.g., crossing minimization heuristics \cite{DBLP:journals/algorithmica/EadesW94,DBLP:journals/tsmc/SugiyamaTT81}, energy-based layout algorithms~\cite{DBLP:journals/congnum/Eades84,DBLP:journals/spe/FruchtermanR91}); for an overview refer to~\cite{DBLP:books/ph/BattistaETT99,DBLP:conf/dagstuhl/1999dg,DBLP:reference/crc/2013gd}.

In this context, Huang et al.~\cite{DBLP:conf/apvis/Huang07,DBLP:journals/vlc/HuangEH14} a decade ago introduced some important experimental evidence,
that edge crossings may not negatively affect the drawing's quality too much (and hence the human's ability to read and interpret it), when the angles formed by the crossing edges are large. In other words, while prior to these experiments it was commonly accepted that mainly the number of crossings is the most important parameter for judging the quality of a non-planar graph drawing, it turned out that the types of edge crossings also matter. As a result, a new and prominent research direction was initiated, recognized under the term ``beyond planarity''~\cite{Shonan2016,Dagstuhl2016,SoCG2017}, which focuses on graphs and their properties, when different constraints on the types of edges crossings are imposed; refer to~\cite{DBLP:journals/corr/abs-1804-07257} for a recent survey.

\begin{figure}[t!]
	\centering
	\subfloat[\label{fig:k5} {}]{
	\includegraphics[page=1,scale=0.85]{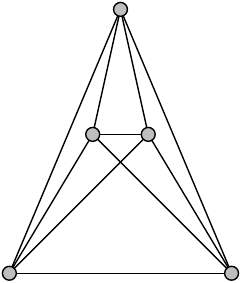}}
	\hfil
	\subfloat[\label{fig:k6} {}]{
	\includegraphics[page=2,scale=0.85]{figures/examples}}
	\caption{%
	(a)~A RAC drawing of the complete graph $K_5$, and
	(b)~a drawing of the complete graph $K_6$, whose crossing resolution is arbitrarily close to $90^\circ$.}
	\label{fig:examples}
\end{figure}

The value of the minimum angle formed by any two crossing edges in a drawing is referred to as its \emph{crossing resolution}; the crossing resolution of a graph is defined as the maximum crossing resolution over all its drawings. Clearly, the crossing resolution of a non-planar graph is at most $90^\circ$, while a graph that admits a drawing with crossing resolution $90^\circ$ is called \emph{right-angle-crossing} (\emph{RAC}) graph; see Figure~\ref{fig:examples}. Notably, RAC graphs are sparse with at most $4n-10$ edges~\cite{DBLP:journals/tcs/DidimoEL11}, while deciding whether a graph is RAC is NP-hard~\cite{DBLP:journals/jgaa/ArgyriouBS12}.

The latter result is an indication that the problem of finding drawings with high crossing resolution might also be difficult, even though, formally, its complexity has not been settled yet for values of the crossing resolution smaller than $90^\circ$. Also, the literature is significantly more limited, when restricting the crossing resolution to be smaller than $90^\circ$, as also evidenced by Section~\ref{sec:relatedwork}.

From a practical point of view, we are only aware of two methods that aim at drawings with high crossing resolution; both of them are adjustments of force-directed algorithms~\cite{DBLP:journals/congnum/Eades84}. The first one is due to Huang et al.~\cite{DBLP:journals/vlc/HuangEHL13}, while the~second one is due to Argyriou et al.~\cite{DBLP:journals/cj/ArgyriouBS13}. Common in both algorithms is that they apply appropriate forces on the endvertices of every pair of crossing edges. Each of them uses a different way to compute (the direction and the magnitude of) the forces, but the underlying idea of both is the same: the smaller the crossing angles are, the larger are the magnitudes of the forces applied at their endvertices.

In this work, we approach the crossing resolution maximization problem from a different perspective. We suggest a simple and intuitive randomization method,
which, in a sense, mimics the way a human would try to increase the crossing resolution of a drawing. How would one increase the crossing resolution of a given drawing? First, she would try to identify the pair of edges that define the crossing resolution of the drawing (we call them \emph{critical} edges); then, she would try to move an endvertex of this pair (which we choose at random), hoping that by this move the crossing resolution will increase. Of course, we cannot consider all possible positions for the vertex to be moved. Instead, we consider a small set of randomly generated ones. If there exists a position among them, that does not lead to a reduction of the crossing resolution, we move the vertex to this~position.

In general, randomization is a technique that has not been deeply examined in Graph Drawing, as it seems difficult to even speculate about the expected quality of the produced drawings; a notable exception is the randomized approach by Goldschmidt and Takvorian~\cite{DBLP:journals/networks/GoldschmidtT94} for computing large planar subgraphs. Since we also could not provide any theoretical guarantee on the expected quality of the produced drawings,
we followed a more practical approach. We implemented our algorithm and the force-directed ones of~\cite{DBLP:journals/cj/ArgyriouBS13} and~\cite{DBLP:journals/vlc/HuangEHL13}, and we experimentally compared them on standard benchmark graphs.
Our evaluation indicates that our method significantly outperforms the force-directed ones~\cite{DBLP:journals/cj/ArgyriouBS13,DBLP:journals/vlc/HuangEHL13} in terms of crossing resolution, but this comes at the cost of slightly worse running time
for large and dense graphs. Analogous results are obtained, when our algorithm and the ones of~\cite{DBLP:journals/cj/ArgyriouBS13} and~\cite{DBLP:journals/vlc/HuangEHL13} are adjusted to maximize  the \emph{angular resolution} (i.e., the minimum value of the angle between any two adjacent edges~\cite{DBLP:journals/siamcomp/FormannHHKLSWW93}) or the \emph{total resolution} (i.e., the minimum of the angular and the crossing resolution~\cite{DBLP:journals/cj/ArgyriouBS13}).

\medskip\noindent{\it Preliminaries:}
Unless otherwise specified, in this paper we consider simple undirected graphs. Let $G=(V,E)$ be such a graph. The degree of vertex $u\in V$ of $G$ is denoted by $d(u)$. The degree $d(G)$ of  graph $G$ is defined as the maximum degree of its vertices, i.e., $d(G)=\max_{u\in V}d(u)$.
Given a drawing $\Gamma(G)$ of $G$, we denote by $p(u)=(x_u,y_u)$ the position of vertex $u \in V$ of $G$ in $\Gamma(G)$. 

\medskip\noindent{\it Structure of the paper:}
The remainder of this paper is structured as follows. Section~\ref{sec:relatedwork} overviews related works.
Our algorithm is presented in detail in Section~\ref{sec:algorithm} and is experimentally evaluated against the ones of Huang et al.~\cite{DBLP:journals/vlc/HuangEHL13} and Argyriou et al.~\cite{DBLP:journals/cj/ArgyriouBS13} in Section~\ref{sec:experiments}, where we also discuss our insights from this project. \arxapp{In the appendix}{In~\cite{arxivVersion}}, we provide experimental results on grid restricted drawings, on more test sets and on the graphs from the Graph Drawing Competition in 2017.

\section{Related Work}
\label{sec:relatedwork}

As already mentioned, the study of the crossing resolution maximization problem has mainly focused on its optimal case, i.e., on the study of RAC graphs. An $n$-vertex RAC graph has at most $4n-10$ edges~\cite{DBLP:journals/tcs/DidimoEL11}, while deciding whether a graph is RAC is NP-hard~\cite{DBLP:journals/jgaa/ArgyriouBS12}. The maximally-dense RAC graphs are 1-planar~\cite{DBLP:journals/dam/EadesL13}, i.e., they can be drawn with at most one crossing per edge. Actually, several~relationships between the class of RAC graphs and subclasses of 1-planar graphs are known~\cite{DBLP:journals/dam/BachmaierBHNR17,DBLP:journals/tcs/BrandenburgDEKL16}. Deciding, however, whether a 1-planar graph is RAC is NP-hard~\cite{DBLP:journals/tcs/BekosDLMM17}. Note that the problem of finding RAC drawings has also been studied in the presence of bends~\cite{DBLP:journals/jgaa/AngeliniCDFBKS11,DBLP:journals/comgeo/ArikushiFKMT12,DBLP:journals/tcs/DidimoEL11,DBLP:journals/mst/GiacomoDLM11} and by imposing restrictions on the degree~\cite{DBLP:conf/s-egc/AngeliniBDFHKLL11}, the structure~\cite{DBLP:journals/ipl/DidimoEL10} and the drawing~\cite{DBLP:journals/algorithmica/GiacomoDEL14,DBLP:conf/wg/HongN15} of the graph. The results are fewer, when the right-angle constraint is relaxed. Dujmovic et al.~\cite{DBLP:journals/cjtcs/DujmovicGMW11} proved that an $n$-vertex graph with crossing resolution at least $\alpha$ radians, has at most $(3n-6)\pi/\alpha$ edges. Corresponding density results are also known in the presence of bends~\cite{DBLP:journals/siamdm/AckermanFT12,DBLP:journals/mst/GiacomoDLM11}.

An immediate observation emerging from the above overview is that the~focus has been primarily on theoretical aspects of the problem. Most of the approaches that could be useful in practice are based on force-directed techniques~\cite{DBLP:books/ph/BattistaETT99,DBLP:journals/congnum/Eades84}. COWA is a system that supports conceptual web site traffic analysis~\cite{DBLP:conf/apvis/DidimoLR10}; its algorithmic core is a force-directed heuristic to compute simultaneous embeddings of two non-planar graphs with high crossing resolution.
Didimo et al.~\cite{DBLP:conf/gd/DidimoLR10} describe topology-driven force-directed heuristics to achieve good trade-offs in terms of number of edge crossings, crossing resolution, and geodesic edge tendency; the obtained drawings, however, are not straight-line.
For straight-line drawings, Nguyen et al.~\cite{DBLP:conf/gd/NguyenEHH10} suggest a quadratic-program to increase the crossing angles of circular drawings.
Of more general scope are the already mentioned force-directed algorithms of Argyriou et al.~\cite{DBLP:journals/cj/ArgyriouBS13} and Huang et al.~\cite{DBLP:journals/vlc/HuangEHL13}.

\section{Description of our Heuristic Approach}
\label{sec:algorithm}

In this section, we describe our heuristic for obtaining drawings with high crossing resolution. The input of our heuristic consists of a graph $G$ and an initial drawing $\Gamma_0$ of $G$ with crossing resolution $c(\Gamma_0)$. We assume that no two edges of~$G$ overlap in $\Gamma_0$, i.e., $c(\Gamma_0)>0$. A circular drawing or a drawing obtained by applying a force-directed algorithm on $G$ clearly meets this precondition.

Our algorithm is iterative and at each iteration performs some operations that are mainly based on randomization. At the $i$-th iteration, we assume that we have computed a drawing $\Gamma_{i-1}$ of crossing resolution $c(\Gamma_{i-1}) \geq c(\Gamma_0)$.
In other words, we assume, as an invariant for our algorithm, that the crossing resolution cannot be decreased at some iteration. Then, a vertex of $\Gamma_{i-1}$ is chosen arbitrarily at random based on the so-called \emph{vertex-pool}, which may contain:
%
\begin{inparaenum}[(i)]
\item either all vertices of $\Gamma_{i-1}$, or
\item a prespecified subset of the vertices of $\Gamma_{i-1}$, called \emph{critical}.
\end{inparaenum}

Intuitively, the critical vertices are the endpoints of the edges that define the crossing resolution of drawing $\Gamma_{i-1}$. To formally define them, we first need to introduce the notion of critical edge-pairs. A pair of edges $e$ and $e'$ is called \emph{critical} in $\Gamma_{i-1}$, if $e$ and $e'$ cross in $\Gamma_{i-1}$ and the minimum angle that is formed at their crossing point is equal to $c(\Gamma_{i-1})$. The set of critical vertices of $\Gamma_{i-1}$ is then defined by the four endvertices of each critical edge-pair.

The role of critical vertices is central in our algorithm\footnote{If the focus is not on the critical vertices for a large graph, then our algorithm will need a large number of iterations to converge to a good solution, because it is simply very unlikely to select to move one of the vertices that define the crossing resolution.}: By appropriately changing the location of a critical vertex or of a vertex in the neighbourhood of the critical vertices, we naturally expect to improve the crossing resolution of the current drawing. We turned this observation into an algorithmic implementation through a probabilistic random selection procedure, so that the vertices at graph-distance~$i$ from the ones of the vertex-pool have higher probability for selection than the corresponding ones at distance $j$  in the graph, when $0 \leq i<j$. So, if the vertex-pool contains only critical vertices, then the closer a vertex is to the critical vertices, the more likely it is to be chosen. Otherwise, the vertex-pool contains all vertices and each vertex can be chosen with the same probability.

What we quickly realized from our practical analysis, is that the crossing resolution of the initial drawing improves rapidly during the first iterations of the algorithm. However, by focusing only at the critical vertices, it is~highly possible that the algorithm will get trapped to some local maxima after a number of iterations. So, special care is needed to avoid these bottlenecks, especially when the input graph is large. We will discuss ways to avoid them later in this section.

So far, we have described the main idea of our algorithm, which at each iteration chooses uniformly at random a vertex of the current drawing to move (based on the content of the vertex-pool), so to improve the crossing resolution. Next, we describe how to compute its new position in the next drawing. Note that our method resembles probabilistic hill climbing approaches.

\begin{figure}[t!]
	\centering
	\subfloat[\label{fig:algo-rays} {}]{
	\includegraphics[page=1, scale=0.9]{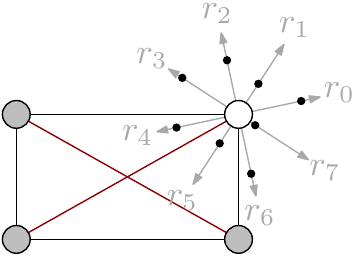}}
	\hfil
	\subfloat[\label{fig:algo-move} {}]{
	\includegraphics[page=2, scale=0.9]{figures/algorithm}}
	\caption{%
	Illustration of an iteration step of our algorithm:
	(a)~The chosen vertex is the white one;
	the computed rays $r_0,\ldots,r_7$ have been rotated by $8^\circ$;
	the black-colored points along these rays are points $\pi_0,\ldots,\pi_7$;
	among them, $\pi_4$ yields the best solution.
	(b)~The resulting drawing after moving the vertex at position $\pi_2$.}
	\label{fig:algo}
\end{figure}

Let $v_i$ be the vertex of $\Gamma_{i-1}$ that has been chosen to be moved at the $i$-th iteration.
To compute the position of $v_i$ in the next drawing $\Gamma_i$, we consider a set of $\rho$ rays $r_0,r_1,\ldots,r_{\rho-1}$ that all emanate from $p(v_i)$ in $\Gamma_{i-1}$, such that the angle formed by ray $r_j$, with $j=0,1,\ldots,\rho-1$, and the horizontal axis equals to $2j\pi/\rho$, where $\rho>0$ is an integer parameter of the algorithm. These rays are then rotated by an angle that is chosen uniformly at random in the interval $[0,2\pi]$; see Fig.~\ref{fig:algo}. The position of vertex $v_i$ in $\Gamma_i$ will eventually be along one of the rays $r_0,r_1,\ldots,r_{\rho-1}$. More precisely, for each ray $r_i$ we choose a distance value $\delta_i$ uniformly at random from the interval $[\delta_{min},\delta_{max}]$, where $\delta_{min}$ and $\delta_{max}$ are two positive parameters of the algorithm. For each $j=0,1,\ldots,\rho-1$, a new point $\pi_j$ is obtained by translating $p(u)$ along $r_j$ by a distance $\delta_j$; point $\pi_j$ is \emph{feasible}, if the crossing resolution of the drawing obtained by placing vertex $v_i$ at $\pi_j$ and by keeping all other vertices of $G$ in their positions in $\Gamma_{i-1}$ is at least as large as the crossing resolution of $\Gamma_{i-1}$, and there is no vertex of $\Gamma_{i-1}$ at $\pi_j$.

If none of the points $\pi_j$, with $j=0,1,\ldots,\rho-1$ is feasible, then the position of $v_i$ in $\Gamma_i$ is $p(v_i)$, i.e., same as in $\Gamma_{i-1}$, since $c(\Gamma_i) \geq c(\Gamma_{i-1})$ must hold. If there is one or more feasible points, then one may consider two different approaches to determine the position of $v_i$ in $\Gamma_i$. The most natural is to choose the feasible point that maximizes the crossing resolution of the obtained drawing. As an alternative, one may rely again on randomization and chose uniformly at random one of the feasible points as the position of $v_i$ in $\Gamma_i$. We note that we did not observe any significant difference between these two approaches (in terms of the crossing resolution of the obtained drawings), so we simply adopted the first one. 
The termination condition of our algorithm is simple and depends on an in\-put parameter~$\tau$. More specifically, if the crossing resolution has not improved~during the last $\tau$ iterations, we assume that the algorithm has converged and we~stop.

\myparagraph{Avoiding local maxima}
%
To avoid getting trapped to locally optimal solutions, we mainly investigated two approaches, which are both  parametrizable by two input parameters $\zeta$ and $\zeta'$. The first mimics the human behaviour. What would~one do to escape from a locally optimal solution? She would stop trying to move the endvertices of the edges defining the crossing resolution; she would rather start moving ``irrelevant'' vertices hoping that
by doing so a better solution will be easier to be computed afterwards. Our algorithm is mimicking this idea as follows:
\begin{inparaenum}[(i)]
\item if during the last $\zeta$ iterations the crossing resolution has not been~improved, then the vertex-pool becomes \emph{wider} by including all the vertices, and the algorithm is executed with this vertex-pool for $\zeta'$ iterations;
\item afterwards, the vertex-pool switches back to the critical vertices.
\end{inparaenum}
%
While this approach turned out to be effective for smaller graphs, for graphs with more than 100 vertices, it was not so efficient; in most iterations with the wider vertex-pool, the embedding could not change in a beneficial way for the algorithm~to~proceed.


Our second approach is based on parameters $\rho$, $\delta_{min}$ and $\delta_{max}$ of the algorithm. Our idea was that if the algorithm gets trapped to a locally optimal solution, then a ``drastic'' or ``sharp'' move may help to escape. We turned this idea into an algorithmic implementation as follows:
\begin{inparaenum}[(i)]
\item if during the last $\zeta$ iterations the crossing resolution has not been improved, we double the values of $\rho$, $\delta_{min}$ and $\delta_{max}$, and the algorithm is executed with these values for $\zeta'$ iterations;
\item afterwards, $\rho$, $\delta_{min}$ and $\delta_{max}$ switch back to this initial value.
\end{inparaenum}
This approach may lead to drawings with larger area, but this is ``expected'', as it turns out that drawings with high crossing resolution may require large area~\cite{DBLP:journals/jgaa/AngeliniCDFBKS11,DBLP:journals/tcs/BrandenburgDEKL16}.

\myparagraph{Complexity issues}
%
A factor that highly affects the efficiency of our algorithm is the computation of the crossing points of the edges and the corresponding angles at these points. Given  a drawing, a na\"ive approach to compute its crossings requires $O(m^2)$ time, which can be improved by a plane-sweep technique to $O(m \log m + c)$ time, where $m$ and $c$ denote the number of edges and crossings.

Instead of computing all crossing points and the corresponding angles for each candidate position of each iteration, we adopted a different approach for determining the set of feasible candidate positions, which turned out to be quite efficient in practice. Recall that we denoted by $v_i$ the vertex chosen at the $i$-th iteration step, and by $\pi_0,\ldots,\pi_{\rho-1}$ the candidate positions to move $v_i$. Let $e_0,\ldots,e_{d_i-1}$ be the edges incident to $v_i$, where $d_i=deg(v_i)$. Next, for each edge $e_k$ with $k=0,\ldots,d_i-1$  we compute the crossings and the corresponding crossing angles of $e_k$ with all other edges in $\Gamma_{i-1}$. Let $\phi_i$ be the minimum crossing angle computed; this is our reference angle. Also, for each candidate position $\pi_j$ with $j=0,\ldots,\rho-1$, and for each edge $e_k$ with $k=0,\ldots,d_i-1$, we compute the crossings and the corresponding crossing angles of $e_k$ with all other edges of the drawing, assuming that $v_i$ is at $\pi_j$. Let $\chi_j$ be the minimum crossing angle computed with this approach, when $v_i$ is at position $\pi_j$. Clearly, $\pi_j$ is feasible only if $\chi_j \geq \phi_i$. Note that the complexity of this approach is $O(deg(v_i)m) = O(nm)$.

\subsection{Some interesting variants}
\label{ssec:variants}

In general, aesthetically pleasant drawings of graphs are usually the result of compromising between different aesthetic criteria. Towards this direction, we discuss in this section interesting variants of our~algorithm, which are motivated by the following observation that we made while working on this project (see Section~\ref{sec:experiments}): Drawings that are optimised only in terms of the crossing resolution tend to have bad aspect ratio
and poor angular resolution.

\myparagraph{Aspect ratio}
%
It was easy to instruct our algorithm to prevent producing drawings with aspect ratio either higher than the one of the starting layout or higher than a given input value. What we simply had to do was to reject candidate positions, which violate this precondition.

\myparagraph{Total resolution}
%
Similarly as above, we could adjust our algorithm to yield drawings with high total resolution by simply taking into account also the angular resolution of the drawing. In particular, if the total resolution of the drawing is defined by its angular resolution, then the way we compute the critical vertices of this drawing has to change; the critical vertices must be the endvertices of the pairs of edges that define the angular resolution. Also, at each iteration of our algorithm we have to ensure that the total resolution does not decrease. We do so by rejecting candidate positions which yield a reduced total~resolution.

\myparagraph{Angular resolution}
%
As it is the case with the force-directed algorithms of Huang et al.~\cite{DBLP:journals/vlc/HuangEHL13} and Argyriou et al.~\cite{DBLP:journals/cj/ArgyriouBS13}, our algorithm can be also restricted to maximize only the angular resolution (by neglecting its crossing resolution). We already described in the previous paragraph the necessary changes in the definition of the critical vertices and the rule according to which a candidate position is rejected (i.e., when it yields a drawing with a reduced angular resolution).

\myparagraph{Grid drawings}
%
Our algorithm, as it has been described so far, does not necessarily produce grid drawings, i.e., drawings in which the vertices are at integer coordinates. However, it can be easily adjusted to produce such drawings. More precisely, if we round the candidate positions computed at each iteration of our algorithm to their closest grid points and use these grid points as candidates for the next position of the vertex to be moved, then the obtained drawing will be grid (assuming, of course, that the starting drawing is grid). One can even bound the size of the grid, by rejecting candidate grid positions outside the bounds. \arxapp{In Appendix~\ref{app:grid}}{In~\cite{arxivVersion}}, we report experimental results on this variant.

\section{Experimental Evaluation}
\label{sec:experiments}

In this section, we present the results of our experimental evaluation. For comparison purposes, apart from our algorithm, we also implemented the force-directed algorithms of Argyriou et al.~\cite{DBLP:journals/cj/ArgyriouBS13} and Huang et al.~\cite{DBLP:journals/vlc/HuangEHL13}. The implementations\footnote{Our implementation is available on request from the authors.} were in Java using yFiles~\cite{DBLP:books/sp/04/WieseE004}. The experiment was performed on a Linux laptop with four cores at 2.4 GHz and 8 GB RAM.
As a test set for our experiment, we used the non-planar Rome graphs~\cite{DBLP:reference/crc/BattistaD13}, which form a collection of around 8.100 benchmark graphs; \arxapp{in Appendix~\ref{app:north}}{in~\cite{arxivVersion}}, we also report on the AT\&T graphs.

\begin{figure}[t]
\centering
\subfloat[\label{fig:cr-res-1}{Crossing resolution vs no.~of vertices}]{
\centering
\scalebox{0.95}{
\centering
\begin{tikzpicture}
\begin{axis}[axis x line*=bottom,xlabel style={yshift=0.2cm},axis y line*=left,ylabel style={yshift=-0.2cm}, legend style={at={(0.425,1.35)},anchor=north,legend columns=2,draw=none}, width=0.475\textwidth, cycle list name=mylist,
mark repeat={5}, xtick={10,20,30,40,50,60,70,80,90,100}, ytick={0,10,20,30,40,50,60,70,80,90}, ylabel={Crossing Resolution}, xlabel={Number of Vertices}, xmin=10, xmax=100, tick pos=left,ymajorgrids]
\addplot table [x=n, y=crossing resolution rm, col sep=semicolon]{crossingResolution.csv};
\addplot table [x=n, y=crossing resolution rm-1, col sep=semicolon] {crossingResolution.csv};
\addplot table [x=n, y=crossing resolution fa, col sep=semicolon]{crossingResolution.csv};
\addplot table [x=Nodes, y=Crossing Only, col sep=semicolon]{abs-crossingResolution.csv};
\legend{{Unrestricted},{AR-restricted~~},{Huang et al.},{Argyriou et al.}}
\end{axis}
\end{tikzpicture}}}
\subfloat[\label{fig:cr-res-2}{Aspect ratio vs no.~of vertices}]{
\centering
\scalebox{0.95}{
\centering
\begin{tikzpicture}
\begin{axis}[axis x line*=bottom,xlabel style={yshift=0.2cm},axis y line*=left,ylabel style={yshift=-0.2cm},legend style={at={(0.425,1.35)}, anchor=north,legend columns=2,draw=none}, width=0.475\textwidth,cycle list name=mylist, mark repeat={5},xtick={10,20,30,40,50,60,70,80,90,100},ylabel={Aspect Ratio}, xlabel={Number of Vertices}, ytick={1,2.5,10,25,100,250,1000,2500,10000}, xmin=10, xmax=100,ymode=log, log ticks with fixed point,tick pos=left, ymajorgrids,]
\addplot table [x=n, y=Aspect ratio rm, col sep=semicolon]{crossingResolution.csv};
\addplot table [x=n, y=Aspect ratio rm-1, col sep=semicolon] {crossingResolution.csv};
\addplot table [x=n, y=Aspect ratio fa, col sep=semicolon]{crossingResolution.csv};
\addplot table [x=Nodes, y=Crossing Only, col sep=semicolon]{abs-aspectRatio.csv};
\legend{{Unrestricted},{AR-restricted~~},{Huang et al.},{Argyriou et al.}}
\end{axis}
\end{tikzpicture}}}

\subfloat[\label{fig:cr-res-3}{No.~of crossings vs no.~of vertices}]{
\centering
\scalebox{0.95}{
\centering
\begin{tikzpicture}
\begin{axis}[axis x line*=bottom,xlabel style={yshift=0.2cm},axis y line*=left,ylabel style={yshift=-0.2cm},legend style={at={(0.425,1.35)}, anchor=north,legend columns=2,draw=none}, width=0.475\textwidth,cycle list name=mylist,  mark repeat={5},xtick={10,20,30,40,50,60,70,80,90,100}, xmin=10, xmax=100, ytick={0,25,50,75,100,125,150,175,200},ylabel={Number of Crossings}, xlabel={Number of Vertices},tick pos=left, ymajorgrids]
\addplot table [x=n, y=Crossing number rm, col sep=semicolon]{crossingResolution.csv};
\addplot table [x=n, y=Crossing number rm-1, col sep=semicolon] {crossingResolution.csv};
\addplot table [x=n, y=Crossing number fa, col sep=semicolon]{crossingResolution.csv};
\addplot table [x=Nodes, y=Crossing Only, col sep=semicolon]{abs-crossings.csv};
\legend{{Unrestricted},{AR-restricted~~},{Huang et al.},{Argyriou et al.}}
\end{axis}
\end{tikzpicture}}}
\subfloat[\label{fig:cr-res-4}{No.~of iterations vs no.~of vertices}]{
\centering
\scalebox{0.95}{
\centering
\begin{tikzpicture}
\begin{axis}[axis x line*=bottom,xlabel style={yshift=0.2cm},axis y line*=left,ylabel style={yshift=-0.2cm},legend style={at={(0.425,1.35)}, anchor=north,legend columns=2,draw=none}, width=0.475\textwidth,cycle list name=mylist,  ylabel={Iterations}, xlabel={Number of Vertices}, mark repeat={5},xtick={10,20,30,40,50,60,70,80,90,100}, xmin=10,
xmax=100, ytick={0,500,1000,1500,2000,2500,3000,3500,4000,4500,5000},tick pos=left, ymajorgrids]
\addplot table [x=n, y=Iterations rm, col sep=semicolon]{crossingResolution.csv};
\addplot table [x=n, y=Iterations rm-1, col sep=semicolon] {crossingResolution.csv};
\addplot table [x=n, y=Iterations fa, col sep=semicolon]{crossingResolution.csv};
\addplot table [x=Nodes, y=Crossing Only, col sep=semicolon]{abs-iterations.csv};
\legend{{Unrestricted},{AR-restricted~~},{Huang et al.},{Argyriou et al.}}
\end{axis}
\end{tikzpicture}}}
\caption{Experimental results on the crossing resolution for the Rome graphs.}
\label{fig:cr-res}
\end{figure}
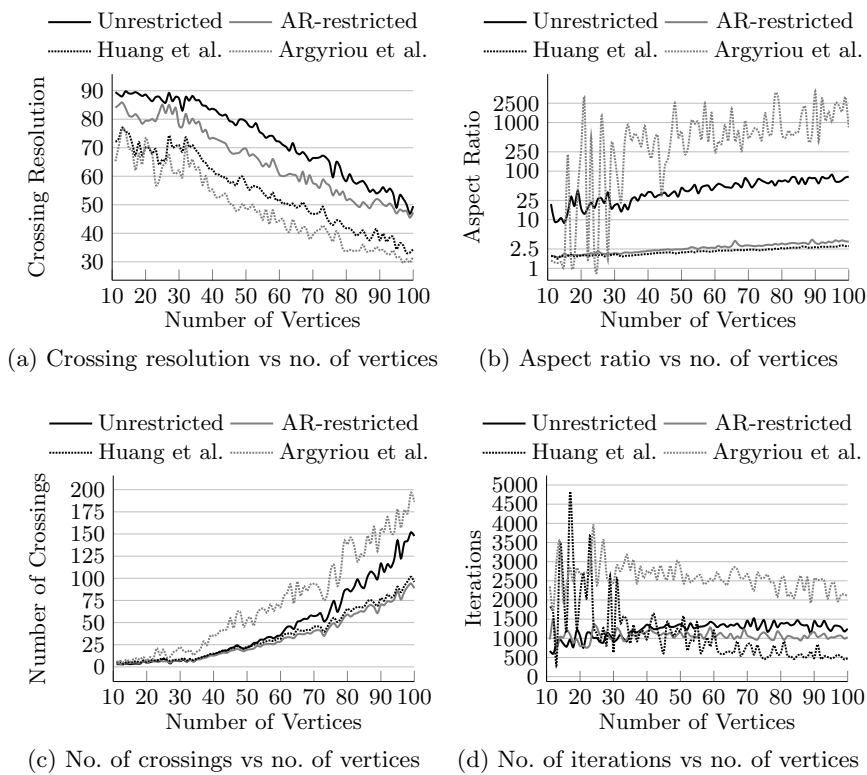

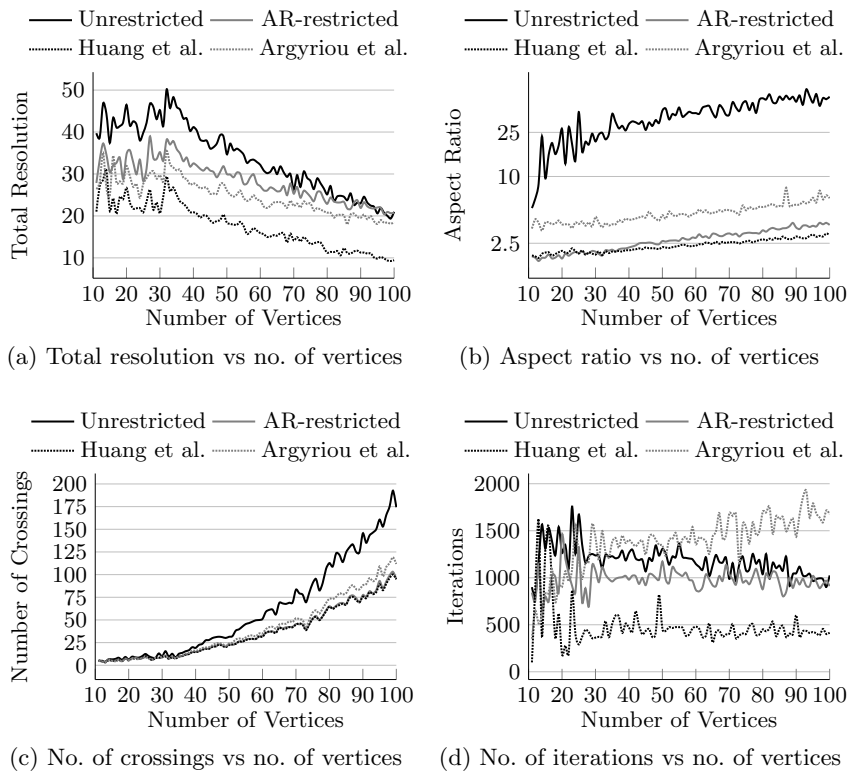
\begin{figure}[t]
\centering
\subfloat[\label{fig:to-res-1}{Total resolution vs no.~of vertices}]{
\centering
\scalebox{0.95}{
\centering
\begin{tikzpicture}
\begin{axis}[axis x line*=bottom,xlabel style={yshift=0.2cm},axis y line*=left,ylabel style={yshift=-0.2cm}, legend style={at={(0.425,1.35)},anchor=north,legend columns=2,draw=none}, width=0.475\textwidth, cycle list name=mylist,
mark repeat={5}, xtick={10,20,30,40,50,60,70,80,90,100}, ytick={0,10,20,30,40,50,60,70,80,90}, ylabel={Total Resolution}, xlabel={Number of Vertices}, xmin=10, xmax=100, tick pos=left,ymajorgrids]
\addplot table [x=n, y=Total Resolution rm, col sep=semicolon]{totalResolution.csv};
\addplot table [x=n, y=Total Resolution rm-1, col sep=semicolon] {totalResolution.csv};
\addplot table [x=n, y=Total Resolution fa, col sep=semicolon]{totalResolution.csv};
\addplot table [x=Nodes, y=Mixed, col sep=semicolon]{abs-totalResolution.csv};
\legend{{Unrestricted},{AR-restricted~~},{Huang et al.},{Argyriou et al.}}
\end{axis}
\end{tikzpicture}}}
\subfloat[\label{fig:to-res-2}{Aspect ratio vs no.~of vertices}]{
\centering
\scalebox{0.95}{
\centering
\begin{tikzpicture}
\begin{axis}[axis x line*=bottom,xlabel style={yshift=0.2cm},axis y line*=left,ylabel style={yshift=-0.2cm},legend style={at={(0.425,1.35)}, anchor=north,legend columns=2,draw=none}, width=0.475\textwidth,cycle list name=mylist, mark repeat={5},xtick={10,20,30,40,50,60,70,80,90,100},ylabel={Aspect Ratio}, xlabel={Number of Vertices}, ytick={1,2.5,10,25,100,250,1000,2500,10000}, xmin=10, xmax=100,ymode=log, log ticks with fixed point,tick pos=left, ymajorgrids,]
\addplot table [x=n, y=Aspect ratio rm, col sep=semicolon]{totalResolution.csv};
\addplot table [x=n, y=Aspect ratio rm-1, col sep=semicolon] {totalResolution.csv};
\addplot table [x=n, y=Aspect ratio fa, col sep=semicolon]{totalResolution.csv};
\addplot table [x=Nodes, y=Mixed, col sep=semicolon]{abs-aspectRatio.csv};
\legend{{Unrestricted},{AR-restricted~~},{Huang et al.},{Argyriou et al.}}
\end{axis}
\end{tikzpicture}}}

\subfloat[\label{fig:to-res-3}{No.~of crossings vs no.~of vertices}]{
\centering
\scalebox{0.95}{
\centering
\begin{tikzpicture}
\begin{axis}[axis x line*=bottom,xlabel style={yshift=0.2cm},axis y line*=left,ylabel style={yshift=-0.2cm},legend style={at={(0.425,1.35)}, anchor=north,legend columns=2,draw=none}, width=0.475\textwidth,cycle list name=mylist,  mark repeat={5},xtick={10,20,30,40,50,60,70,80,90,100}, xmin=10, xmax=100, ytick={0,25,50,75,100,125,150,175,200},ylabel={Number of Crossings}, xlabel={Number of Vertices},tick pos=left, ymajorgrids]
\addplot table [x=n, y=Crossing number rm, col sep=semicolon]{totalResolution.csv};
\addplot table [x=n, y=Crossing number rm-1, col sep=semicolon] {totalResolution.csv};
\addplot table [x=n, y=Crossing number fa, col sep=semicolon]{totalResolution.csv};
\addplot table [x=Nodes, y=Mixed, col sep=semicolon]{abs-crossings.csv};
\legend{{Unrestricted},{AR-restricted~~},{Huang et al.},{Argyriou et al.}}
\end{axis}
\end{tikzpicture}}}
\subfloat[\label{fig:to-res-4}{No.~of iterations vs no.~of vertices}]{
\centering
\scalebox{0.95}{
\centering
\begin{tikzpicture}
\begin{axis}[axis x line*=bottom,xlabel style={yshift=0.2cm},axis y line*=left,ylabel style={yshift=-0.2cm},legend style={at={(0.425,1.35)}, anchor=north,legend columns=2,draw=none}, width=0.475\textwidth,cycle list name=mylist,  ylabel={Iterations}, xlabel={Number of Vertices}, mark repeat={5},xtick={10,20,30,40,50,60,70,80,90,100}, xmin=10,
xmax=100, ytick={0,500,1000,1500,2000,2500,3000,3500,4000,4500,5000},tick pos=left, ymajorgrids]
\addplot table [x=n, y=Iterations rm, col sep=semicolon]{totalResolution.csv};
\addplot table [x=n, y=Iterations rm-1, col sep=semicolon] {totalResolution.csv};
\addplot table [x=n, y=Iterations fa, col sep=semicolon]{totalResolution.csv};
\addplot table [x=Nodes, y=Mixed, col sep=semicolon]{abs-iterations.csv};
\legend{{Unrestricted},{AR-restricted~~},{Huang et al.},{Argyriou et al.}}
\end{axis}
\end{tikzpicture}}}
\caption{Experimental results on the total resolution for the Rome graphs.}
\label{fig:to-res}
\end{figure}

\begin{figure}[t]
\centering
\subfloat[\label{fig:an-res-1}{Angular resolution vs no.~of vertices}]{
\centering
\scalebox{0.95}{
\centering
\begin{tikzpicture}
\begin{axis}[axis x line*=bottom,xlabel style={yshift=0.2cm},axis y line*=left,ylabel style={yshift=-0.2cm}, legend style={at={(0.425,1.35)},anchor=north,legend columns=2,draw=none}, width=0.475\textwidth, cycle list name=mylist,
mark repeat={5}, xtick={10,20,30,40,50,60,70,80,90,100}, ytick={0,10,20,30,40,50,60,70,80,90}, ylabel={Angular Resolution}, xlabel={Number of Vertices}, xmin=10, xmax=100, tick pos=left,ymajorgrids]
\addplot table [x=n, y=Angular resolution rm, col sep=semicolon]{angularResolution.csv};
\addplot table [x=n, y=Angular resolution rm-1, col sep=semicolon] {angularResolution.csv};
\addplot table [x=n, y=Angular resolution fa, col sep=semicolon]{angularResolution.csv};
\addplot table [x=Nodes, y=Angular Only, col sep=semicolon]{abs-angularResolution.csv};
\legend{{Unrestricted},{AR-restricted~~},{Huang et al.},{Argyriou et al.}}
\end{axis}
\end{tikzpicture}}}
\subfloat[\label{fig:an-res-2}{Aspect ratio vs no.~of vertices}]{
\centering
\scalebox{0.95}{
\centering
\begin{tikzpicture}
\begin{axis}[axis x line*=bottom,xlabel style={yshift=0.2cm},axis y line*=left,ylabel style={yshift=-0.2cm},legend style={at={(0.425,1.35)}, anchor=north,legend columns=2,draw=none}, width=0.475\textwidth,cycle list name=mylist, mark repeat={5},xtick={10,20,30,40,50,60,70,80,90,100},ylabel={Aspect Ratio}, xlabel={Number of Vertices}, ytick={1,2.5,10,25,100,250,1000,2500,10000}, xmin=10, xmax=100,ymode=log, log ticks with fixed point,tick pos=left, ymajorgrids,]
\addplot table [x=n, y=Aspect ratio rm, col sep=semicolon]{angularResolution.csv};
\addplot table [x=n, y=Aspect ratio rm-1, col sep=semicolon] {angularResolution.csv};
\addplot table [x=n, y=Aspect ratio fa, col sep=semicolon]{angularResolution.csv};
\addplot table [x=Nodes, y=Angular Only, col sep=semicolon]{abs-aspectRatio.csv};
\legend{{Unrestricted},{AR-restricted~~},{Huang et al.},{Argyriou et al.}}
\end{axis}
\end{tikzpicture}}}

\subfloat[\label{fig:an-res-3}{No.~of crossings vs no.~of vertices}]{
\centering
\scalebox{0.95}{
\centering
\begin{tikzpicture}
\begin{axis}[axis x line*=bottom,xlabel style={yshift=0.2cm},axis y line*=left,ylabel style={yshift=-0.2cm},legend style={at={(0.425,1.35)}, anchor=north,legend columns=2,draw=none}, width=0.475\textwidth,cycle list name=mylist,  mark repeat={5},xtick={10,20,30,40,50,60,70,80,90,100}, xmin=10, xmax=100, ytick={0,25,50,75,100,125,150,175,200},ylabel={Number of Crossings}, xlabel={Number of Vertices},tick pos=left, ymajorgrids]
\addplot table [x=n, y=Crossing number rm, col sep=semicolon]{angularResolution.csv};
\addplot table [x=n, y=Crossing number rm-1, col sep=semicolon] {angularResolution.csv};
\addplot table [x=n, y=Crossing number fa, col sep=semicolon]{angularResolution.csv};
\addplot table [x=Nodes, y=Angular Only, col sep=semicolon]{abs-crossings.csv};
\legend{{Unrestricted},{AR-restricted~~},{Huang et al.},{Argyriou et al.}}
\end{axis}
\end{tikzpicture}}}
\subfloat[\label{fig:an-res-4}{No.~of iterations vs no.~of vertices}]{
\centering
\scalebox{0.95}{
\centering
\begin{tikzpicture}
\begin{axis}[axis x line*=bottom,xlabel style={yshift=0.2cm},axis y line*=left,ylabel style={yshift=-0.2cm},legend style={at={(0.425,1.35)}, anchor=north,legend columns=2,draw=none}, width=0.475\textwidth,cycle list name=mylist,  ylabel={Iterations}, xlabel={Number of Vertices}, mark repeat={5},xtick={10,20,30,40,50,60,70,80,90,100}, xmin=10,
xmax=100, ytick={0,500,1000,1500,2000,2500,3000,3500,4000,4500,5000},tick pos=left, ymajorgrids]
\addplot table [x=n, y=Iterations rm, col sep=semicolon]{angularResolution.csv};
\addplot table [x=n, y=Iterations rm-1, col sep=semicolon] {angularResolution.csv};
\addplot table [x=n, y=Iterations fa, col sep=semicolon]{angularResolution.csv};
\addplot table [x=Nodes, y=Angular Only, col sep=semicolon]{abs-iterations.csv};
\legend{{Unrestricted},{AR-restricted~~},{Huang et al.},{Argyriou et al.}}
\end{axis}
\end{tikzpicture}}}
\caption{Experimental results on the angular resolution for the Rome graphs.}
\label{fig:an-res}
\end{figure}
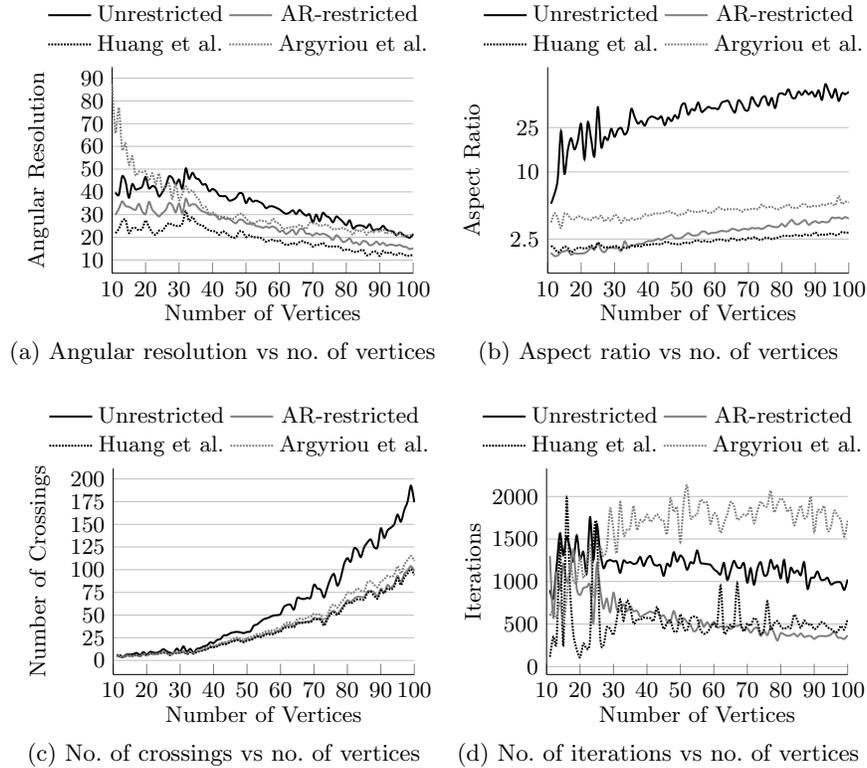

The experiment was performed as follows. Initially, each Rome graph was laid out using the SmartOrganic layouter of yFiles~\cite{DBLP:books/sp/04/WieseE004}. Starting from this layout, every graph was drawn with
\begin{inparaenum}[(i)]
\item our algorithm,
\item our algorithm restricted~not to violate the aspect ratio of the initial layout, and the force-directed algorithms
\item by Argyriou et al.\ and
\item by Huang et al.
\end{inparaenum}
%
%
Since all algorithms of the experiment can easily be adjusted to maximize~only the crossing resolution, or only the angular resolution or both (by maximizing~the total resolution), we adjusted each of the algorithms to maximize exclusively the corresponding measures; see Figs.~\ref{fig:cr-res}, \ref{fig:to-res} and~\ref{fig:an-res}. In our algorithm, this can be achieved by modifying appropriately the content of the vertex-pool (as we saw in Section~\ref{ssec:variants}), while in the algorithms of Argyriou et al.\ and of Huang et al.\ by switching on only the forces that maximize the corresponding properties under measure (note that, each of these two algorithms has a different set of forces to maximize the crossing and the angular resolution, such that together they maximize the total resolution). The reported results are on average across different drawings with same number of vertices. Finally, we mention that for our algorithm, we chose $\delta_{max}=\frac{1}{2}\max\{w,h\}$, where $w$ and $h$ are the width and the height of the initial drawing, respectively, $\delta_{min}=\frac{1}{100}\delta_{max}$ and $\rho=10$. 


\myparagraph{Crossing resolution} Our results for the crossing resolution are summarized in Fig.~\ref{fig:cr-res}. Here, each algorithm was adjusted to maximize exclusively the crossing resolution (i.e., by ignoring the drawing's angular resolution). It is immediate to see that our algorithm outperforms all other ones in terms of the crossing resolution of the produced drawings, when we do not impose any restriction on the aspect ratio of the computed drawings; refer to the solid-black curve, denoted as \emph{Unrestricted}, in Fig.~\ref{fig:cr-res-1}. The variant of our algorithm, which does not violate the aspect ratio of the initial layout, leads to drawings with slightly smaller crossing resolution; refer to the solid-gray curve, denoted as \emph{AR-restricted}, in Fig.~\ref{fig:cr-res-1}. Finally, the two force-directed algorithms seem to produce drawings with worse crossing resolution; refer to the dotted-gray and dotted-black curves of Fig.~\ref{fig:cr-res-1} (by Argyriou et al.\ and by Huang et al., respectively).

While our unrestricted algorithm produces drawings with better crossing resolution, this comes at a cost of drastically increased aspect ratio (see Fig.~\ref{fig:cr-res-2}), which, however, is still better that the corresponding aspect ratio of the drawings produced by the algorithm of Argyriou et al. For the latter algorithm, it seems that the forces due to the angles formed at the crossings outperform the corresponding spring forces, which try to keep the lengths of the edges short. Going back to our unrestricted algorithm, its behaviour is up to a certain degree expected, mainly due to the fact that there is no control on the lengths of the edges.
On the other hand, the restricted variant of our algorithm, which does not allow the aspect ratio to increase, has more or less comparable performance (in terms of aspect ratio) with the one of Huang et al.

Regarding the number of crossings, the restricted variant of our algorithm and the force-directed algorithm of Huang et al.\ yield drawings with comparable number of crossings, which at the same time is significantly smaller than the  number of crossings produced by the two other algorithms; see Fig.~\ref{fig:cr-res-3}.

A different behaviour can be observed in the number of iterations, which are required by the algorithms to converge; refer to Fig.~\ref{fig:cr-res-4}. We note here that we used different criteria to determine whether the algorithms of our experiment had converged. For our algorithms and for the force-directed algorithm by Huang et al., we assumed that the algorithm had converged, if the crossing resolution between 500 consecutive iterations was not improved by more than 0.001 degrees. For the algorithm by Argyriou et al., we decided to use a much more restricted convergence criterion, because the produced layouts can change vastly between  consecutive iterations. We made this choice mainly to have ``comparable'' number of iterations among the algorithms of the experiment. In this direction, we adopted the convergence criterion that the authors used in their previous experimental analysis\, that is, we assumed that the algorithm had converged, if the crossing resolution between two consecutive iterations was not improved by more than 0.001 degrees. Observe that even under this more restricted convergence criterion, the algorithm needs significantly more iterations to converge than the remaining three algorithms of the experiment; see Fig.~\ref{fig:cr-res-4}. The maximum number of iterations that each of the algorithms could perform in order to converge was set to 100.000, but that limit was never reached. We observe that both force-directed algorithms seem to require a great amount of iterations to converge for small graphs, where a drawing with really good crossing resolution is possible. However, for larger graphs  the algorithm by Huang et al.\ requires the least amount of iterations. On the other hand, both the unrestricted and the restricted variant of our algorithm require comparable number of iterations to converge, but clearly more than the ones of the algorithm by Huang et al.

\myparagraph{Total resolution} Our results for the total resolution are summarized in Fig.~\ref{fig:to-res}. Here, each algorithm was adjusted to maximize both the crossing and the angular resolution. For the vast majority of the graphs in the experiment, both our unrestricted algorithm and its restricted variant yield  drawings with better total resolution than the corresponding ones by Argyriou et al. The drawings produced by the algorithm by Huang et al.\ seems to have worse total resolution; see Fig.~\ref{fig:to-res-1}. Note, however, that both variants of our algorithm as well as the force-directed algorithm by Argyriou et al.\ tend to produce drawings of the same total resolution for larger graphs with a small difference in our favor.

Contrary to the results for the total resolution, the results for the aspect ratio show that the drawings produced by the algorithm by Huang et al.\ are better (in terms of aspect ratio) than the drawings produced by remaining algorithms; see Fig.~\ref{fig:to-res-2}. More concretely, the drawings produced by the restricted variant of our algorithm have slightly worse aspect ratios. Then, the ones produced by the force-directed algorithm by Argyriou et al.\ follow. Again, we observe that our unrestricted algorithm leads to drawings with very high aspect ratio.

The restricted variant of our algorithm and the algorithm by Huang et al.\ yield drawings with the least number of crossings; see Fig.~\ref{fig:to-res-3}. Comparable but slightly worse (in terms of the number of crossings) are the drawings produced by the force-directed algorithm by Argyriou et al. Our unrestricted algorithm seems to require the largest number of crossings, which turn out to be notably higher than the corresponding ones of the other three algorithms.

On the negative side, both the unrestricted and the restricted variant of our algorithm require more iterations than the  force-directed algorithm by Huang et al.; see Fig.~\ref{fig:to-res-4}. Recall, however, that the latter algorithm is clearly outperformed by both our variants in term of total resolution. The algorithm by Argyriou et al.\ clearly requires the highest number of iterations (especially for large graphs). We note that the convergence criterion was the same as for the crossing resolution; however, the measured quality was (not the crossing but) the total resolution.

\myparagraph{Angular resolution} We conclude the analysis of our experimental evaluation with the  results for the angular resolution; see Fig.~\ref{fig:an-res}. Here, each algorithm was adjusted  to maximize only the angular resolution (i.e., by ignoring the drawing's crossing resolution). A notable observation is that, for small graphs the best results are achieved by the algorithm by Argyriou et al., while for medium-size graphs by our unrestricted algorithm; see Fig.~\ref{fig:an-res-1}. For large graphs, the two algorithms tend to have the same performance. The restricted variant of our algorithm yields drawings with slightly worse angular resolution. The algorithm by Huang et al.\ is outperformed by all algorithms of the experiment.

The results for the aspect ratio, the number of crossings and the required number of iterations are very similar with corresponding ones for the total resolution; see Figs.~\ref{fig:an-res-2}--\ref{fig:an-res-4}. This observation suggests that, for most of the graphs of our experiment, the angular resolution dominates the crossing resolution (and thus is the one defining the total resolution) in the constructed drawings, which explains the similarity in the reported results. The small differences result from the fact that the crossing resolution cannot be entirely neglected.

\myparagraph{Discussion}
%
%
%
While working on this project, we made some useful observations and  obtained some interesting insights. In particular, there is a recent hypothesis (also supported by experiments) that drawings, in which the crossing angles, are large are easy to read and understand. We observed that drawings that are optimized only in terms of the crossing angles might be arbitrarily bad and may have several undesired properties. In particular, in these drawings it was very common to have adjacent edges to run almost in parallel and vertices to be very close to each other. Hence, angular resolution and aspect ratio were often poor. The additional restrictions that we imposed regarding the angular resolution and the aspect ratio helped significantly improving the readability of the drawings, without loosing too much of their quality in terms of the crossing resolution.

We conclude by noting that our motivation to work with this problem was our participation to GD2017 contest, where we performed miserably using a force-directed algorithm; for details \arxapp{see Appendix~\ref{app:contest}}{see~\cite{arxivVersion}}. As our evaluation shows, the performance of such algorithms is good, only when several aesthetic criteria are taken into account; our new approach is definitely more promising than our previous as evidenced by our experiments. The framework that we developed seems to be quite adaptable to optimize or to take into account also other desired aesthetic properties of a drawing.

\paragraph{Acknowlegdments.} This project was supported by DFG grant KA812/18-1. The authors would also like to acknowledge Simon Wegendt and Jessica Wolz for implementing the first version of our prototype.

\bibliographystyle{abbrvurl}
\bibliography{references}

\arxapp{\newpage
\appendix

\section*{Appendix}

\section{Experiments on Grid Drawings}
\label{app:grid}

\begin{figure}[!b]
\centering
\subfloat[\label{fig:crossingResolutionGrid}{Crossing resolution vs no.~of vertices}]{
\centering
\scalebox{0.95}{
\centering
\begin{tikzpicture}
\begin{axis}[
axis x line*=bottom,xlabel style={yshift=0.2cm},ylabel style={yshift=-0.2cm},axis y line*=left,
legend style={at={(0.425,1.35)},anchor=north,,legend columns=2,draw=none},
width=0.475\textwidth,
cycle list name=mylist,
mark repeat={5},xlabel={Number of Vertices},ylabel={Crossing Resolution},
xtick={10,20,30,40,50,60,70,80,90,100},
xmin=10,
xmax=100,
ytick={10,20,30,40,50,60,70,80,90},
tick pos=left,ymajorgrids]
\addplot table [x=n, y=crossing resolution rm1M, col sep=semicolon]{gridExperiment.csv};
\addplot table [x=n, y=crossing resolution rm10k, col sep=semicolon] {gridExperiment.csv};
\addplot table [x=n, y=crossing resolution rm1000, col sep=semicolon]{gridExperiment.csv};
\addplot table [x=n, y=crossing resolution rm100, col sep=semicolon] {gridExperiment.csv};
\addplot table [x=n, y=crossing resolution rm, col sep=semicolon]{gridExperiment.csv};
\legend{{$10^6 \times 10^6$},{$10^4 \times 10^4$},{$10^3 \times 10^3$},{$10^2 \times 10^2$}}
\end{axis}
\end{tikzpicture}}}
\subfloat[\label{fig:crossingResolutionAspectRatioGrid}{Aspect ratio vs no.~of vertices}]{
\centering
\scalebox{0.95}{
\centering
\begin{tikzpicture}
\begin{axis}[axis x line*=bottom,xlabel style={yshift=0.2cm},axis y line*=left,legend style={at={(0.425,1.35)},
anchor=north,legend columns=2,draw=none},ylabel style={yshift=-0.2cm}, width=0.475\textwidth,cycle list name=mylist, mark repeat={5},xtick={0,10,20,30,40,50,60,70,80,90,100},ylabel={Aspect Ratio},
xlabel={Number of Vertices},
xmin=10,
xmax=100,
ytick={1,2,5,10,20,50,100,200},ymode=log, log ticks with fixed point,tick pos=left, ymajorgrids,]
\addplot table [x=n, y=Aspect ratio rm1M, col sep=semicolon]{gridExperiment.csv};
\addplot table [x=n, y=Aspect ratio rm10k, col sep=semicolon] {gridExperiment.csv};
\addplot table [x=n, y=Aspect ratio rm1000, col sep=semicolon]{gridExperiment.csv};
\addplot table [x=n, y=Aspect ratio rm100, col sep=semicolon] {gridExperiment.csv};
\addplot table [x=n, y=Aspect ratio rm, col sep=semicolon]{gridExperiment.csv};
\legend{{$10^6 \times 10^6$},{$10^4 \times 10^4$},{$10^3 \times 10^3$},{$10^2 \times 10^2$}}
\end{axis}
\end{tikzpicture}}}

\subfloat[\label{fig:crossingResolutionCrossingNumberGrid}{No.~of crossings vs no.~of vertices}]{
\centering
\scalebox{0.95}{
\centering
\begin{tikzpicture}
\begin{axis}[axis x line*=bottom,xlabel style={yshift=0.2cm},axis y line*=left,legend style={at={(0.425,1.35)},
anchor=north,legend columns=2,draw=none}, width=0.475\textwidth,cycle list name=mylist,  mark repeat={5},xtick={10,20,30,40,50,60,70,80,90,100},
xmin=10,
xmax=100,
ytick={0,250,500,750,1000,1250,1500,1750},ylabel={Number of Crossings},
xlabel={Number of Vertices},tick pos=left,ylabel style={yshift=-0.2cm}, ymajorgrids]
\addplot table [x=n, y=Crossing number rm1M, col sep=semicolon]{gridExperiment.csv};
\addplot table [x=n, y=Crossing number rm10k, col sep=semicolon] {gridExperiment.csv};
\addplot table [x=n, y=Crossing number rm1000, col sep=semicolon]{gridExperiment.csv};
\addplot table [x=n, y=Crossing number rm100, col sep=semicolon] {gridExperiment.csv};
\addplot table [x=n, y=Crossing number rm, col sep=semicolon] {gridExperiment.csv};
\legend{{$10^6 \times 10^6$},{$10^4 \times 10^4$},{$10^3 \times 10^3$},{$10^2 \times 10^2$}}
\end{axis}
\end{tikzpicture}}}
\subfloat[\label{fig:crossingResolutionTimeGrid}{Iterations vs no.~of vertices}]{
\centering
\scalebox{0.95}{
\centering
\begin{tikzpicture}
\begin{axis}[axis x line*=bottom,xlabel style={yshift=0.2cm},ylabel style={yshift=-0.2cm},axis y line*=left,legend style={at={(0.425,1.35)},
anchor=north,legend columns=2,draw=none}, width=0.475\textwidth,cycle list name=mylist, mark repeat={5},xtick={10,20,30,40,50,60,70,80,90,100},
xmin=10,
xmax=100,
ytick={0,1000,2000,3000,4000,5000,6000,7000,8000},tick pos=left,
ylabel={Iterations},
xlabel={Number of Vertices},ymajorgrids]
\addplot table [x=n, y=Iterations rm1M, col sep=semicolon]{gridExperiment.csv};
\addplot table [x=n, y=Iterations rm10k, col sep=semicolon] {gridExperiment.csv};
\addplot table [x=n, y=Iterations rm1000, col sep=semicolon]{gridExperiment.csv};
\addplot table [x=n, y=Iterations rm100, col sep=semicolon] {gridExperiment.csv};
\addplot table [x=n, y=Iterations rm, col sep=semicolon] {gridExperiment.csv};
\legend{{$10^6 \times 10^6$},{$10^4 \times 10^4$},{$10^3 \times 10^3$},{$10^2 \times 10^2$}}
\end{axis}
\end{tikzpicture}}}
\caption{Our experimental results on the crossing resolution with different grid restrictions.
The double line corresponds to our unrestricted algorithm.}
\label{fig:experimentsGrid}
\end{figure}
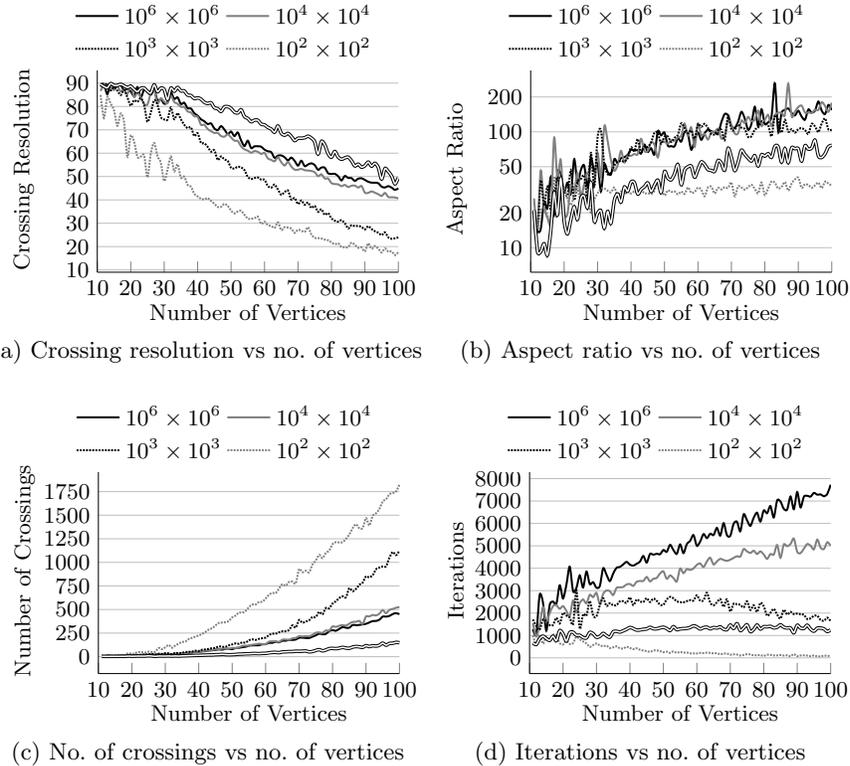

In addition to the experiments described in Section~\ref{sec:experiments}, we also evaluated how our algorithm performs, if we restrict its vertices to integer grid coordinates. In particular, we were interesting to see how the different quality measures that we evaluated in Section~\ref{sec:experiments} are affected by the restrictions imposed on having the vertices of the graph on integer grids of different sizes:
\begin{inparaenum}[(i)]
\item $10^6 \times 10^6$
\item $10^4 \times 10^4$
\item $10^3 \times 10^3$, and
\item $10^2 \times 10^2$.
\end{inparaenum}
The test suite for this experiment was again the non-planar Rome graphs~\cite{DBLP:reference/crc/BattistaD13}. However, since our algorithm is guaranteed to produce a grid drawing, only if its initial drawing is grid, each of the Rome graphs was initially laid out by randomly placing its vertices on the grid, ensuring that neither two vertices nor two edges overlap. The layouts for each of the different sizes of the grid were computed with the variant of our algorithm that optimizes the crossing resolution; Fig.~\ref{fig:experimentsGrid} summarizes the results.

Regarding the crossing resolution, we can observe that with increasing grid size, we could achieve better crossing resolution; see Fig.~\ref{fig:crossingResolutionGrid}. More precisely, a grid of size $10^2 \times 10^2$ was too restrictive for the vast majority of the graphs. As a result, the reported drawings were often the initial ones (as our algorithm could not improve them), especially for large graphs. Significantly fewer were the graphs for which our algorithm could not report an improved drawing, when the grid size was set to $10^3 \times 10^3$.
For grid size $10^4 \times 10^4$, the drawings produced by our algorithm were on average by only $10^\circ$ worse than those produced by the unrestricted version of our algorithm (double line in Fig.~\ref{fig:crossingResolutionGrid}), while the gap was closer for grid size $10^6 \times 10^6$.

The aspect ratio of the computed drawings was more or less the same regardless of the size of the underlying grid, with the exception of the drawings computed on the grid of size $10^2 \times 10^2$; see Fig.~\ref{fig:crossingResolutionAspectRatioGrid}. The fact that the aspect ratio of these drawings was worse can be explained of course by the fact that in most cases an improved drawing could not be reported.

As expected, the smaller the underlying grid is, the more crossings the computed drawings contain; see Fig.~\ref{fig:crossingResolutionCrossingNumberGrid}. As a result, the unrestricted variant of our algorithm clearly outperforms all other ones. It is worth noting that the differences are clear between grid sizes $10^2 \times 10^2$, $10^3 \times 10^3$ and $10^4 \times 10^4$. Notably, there is only a slight improvement (in terms of the number of crossings) from grid size $10^4 \times 10^4$ to $10^6 \times 10^6$. 
On the other hand, the number of iterations needed for convergence increases with the grid size (see Fig.~\ref{fig:crossingResolutionTimeGrid}), with the exception of the grid of size $10^2 \times 10^2$, which verifies our previous observation that for the vast majority of the graphs an improved drawing could not be reported.

In conclusion, we can state that our algorithm is still able to compute drawings with high crossing resolution when restricted to a grid, as long as the grid is not too small. However, the computation of a grid drawing takes longer, which is of course expected.
Finally, note that the choice of the initial grid drawing seems to affect the performance of our algorithm, both with respect to the quality of the produced drawings (counted here in terms of the crossing resolution) but also with respect to the number of iterations needed to converge.

\section{Experiments on the AT\&T  Graph Test Set}
\label{app:north}

In this section, we report the results of our expertimental evaluation (on the crossing, total and angular resolutions) for the non-planar AT\&T graphs, which form a collection of 424 benchmark graphs (also known as Graph Catalog and North graphs; available at \texttt{http://graphdrawing.org/data}). Note that we did not impose any grid constraint on our algorithms. The corresponding results are illustrated in Figs.~\ref{fig:northCrossing}, \ref{fig:northTotal} and~\ref{fig:northAngular}. In general, we observed that the variance of the results is much larger than in the experiments on the Rome graphs. This manifests in spikes of large magnitude in the illustrations of the results and indicates that the structural properties of the graphs in this second test set varies vastly between different graph sizes.

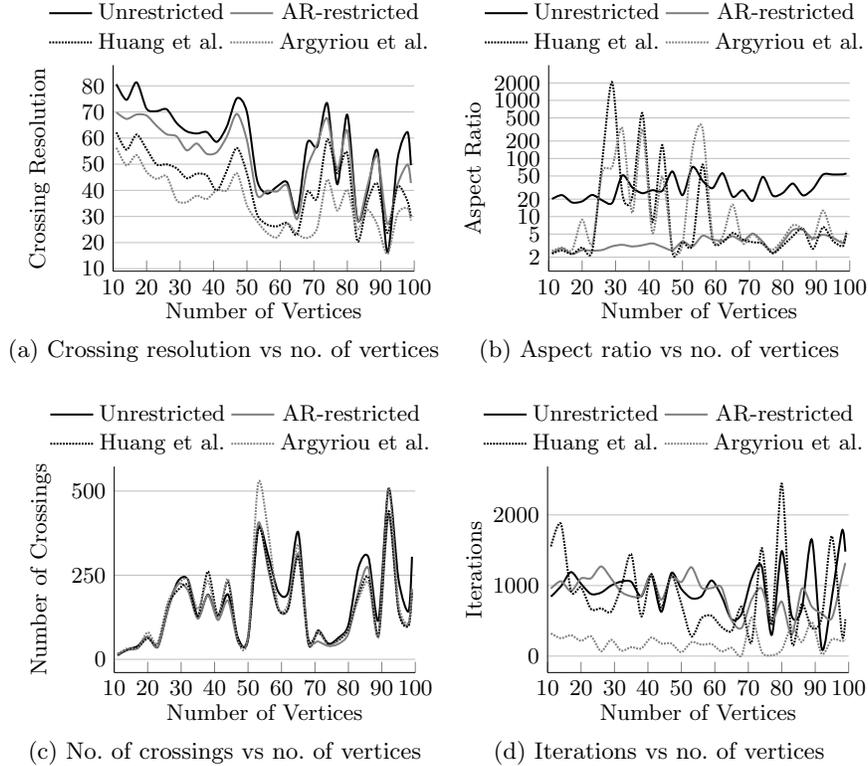
\begin{figure}[htb]
\centering
\subfloat[\label{fig:northCrossing-1}{Crossing resolution vs no.~of vertices}]{
\centering
\scalebox{.95}{
\centering
\begin{tikzpicture}
\begin{axis}[
axis x line*=bottom,xlabel style={yshift=0.2cm},ylabel style={yshift=-0.2cm},axis y line*=left,
legend style={at={(0.425,1.35)},anchor=north,,legend columns=2,draw=none},
width=0.475\textwidth,
cycle list name=mylist,
mark repeat={5},xlabel={Number of Vertices},ylabel={Crossing Resolution},
xtick={10,20,30,40,50,60,70,80,90,100},
xmin=10,
xmax=100,
ytick={10,20,30,40,50,60,70,80,90},
tick pos=left,ymajorgrids]
\addplot table [x=avg, y=cr rm, col sep=semicolon]{crossingResolutionNorth.csv};
\addplot table [x=avg, y=cr rm-1, col sep=semicolon] {crossingResolutionNorth.csv};
\addplot table [x=avg, y=cr fa, col sep=semicolon]{crossingResolutionNorth.csv};
\addplot table [x=avg, y=cr tr, col sep=semicolon] {crossingResolutionNorth.csv};
\legend{{Unrestricted},{AR-restricted~~},{Huang et al.},{Argyriou et al.}}
\end{axis}
\end{tikzpicture}}}
\subfloat[\label{fig:northCrossing-2}{Aspect ratio vs no.~of vertices}]{
\centering
\scalebox{.95}{
\centering
\begin{tikzpicture}
\begin{axis}[axis x line*=bottom,xlabel style={yshift=0.2cm},axis y line*=left,legend style={at={(0.425,1.35)},
anchor=north,legend columns=2,draw=none},ylabel style={yshift=-0.2cm}, width=0.475\textwidth,cycle list name=mylist, mark repeat={5},xtick={0,10,20,30,40,50,60,70,80,90,100},ylabel={Aspect Ratio},
xlabel={Number of Vertices},
xmin=10,
xmax=100,
ytick={1,2,5,10,20,50,100,200,500,1000,2000},ymode=log, log ticks with fixed point,tick pos=left, ymajorgrids,]
\addplot table [x=avg, y=ar rm, col sep=semicolon]{crossingResolutionNorth.csv};
\addplot table [x=avg, y=ar rm-1, col sep=semicolon] {crossingResolutionNorth.csv};
\addplot table [x=avg, y=ar fa, col sep=semicolon]{crossingResolutionNorth.csv};
\addplot table [x=avg, y=ar tr, col sep=semicolon] {crossingResolutionNorth.csv};
\legend{{Unrestricted},{AR-restricted~~},{Huang et al.},{Argyriou et al.}}
\end{axis}
\end{tikzpicture}}}

\subfloat[\label{fig:northCrossing-3}{No.~of crossings vs no.~of vertices}]{
\centering
\scalebox{.95}{
\centering
\begin{tikzpicture}
\begin{axis}[axis x line*=bottom,xlabel style={yshift=0.2cm},axis y line*=left,legend style={at={(0.425,1.35)},
anchor=north,legend columns=2,draw=none}, width=0.475\textwidth,cycle list name=mylist,  mark repeat={5},xtick={10,20,30,40,50,60,70,80,90,100},
xmin=10,
xmax=100,
ytick={0,250,500,750,1000,1250,1500,1750},ylabel={Number of Crossings},
xlabel={Number of Vertices},tick pos=left,ylabel style={yshift=-0.2cm}, ymajorgrids]
\addplot table [x=avg, y=cn rm, col sep=semicolon]{crossingResolutionNorth.csv};
\addplot table [x=avg, y=cn rm-1, col sep=semicolon] {crossingResolutionNorth.csv};
\addplot table [x=avg, y=cn fa, col sep=semicolon]{crossingResolutionNorth.csv};
\addplot table [x=avg, y=cn tr, col sep=semicolon] {crossingResolutionNorth.csv};
\legend{{Unrestricted},{AR-restricted~~},{Huang et al.},{Argyriou et al.}}
\end{axis}
\end{tikzpicture}}}
\subfloat[\label{fig:northCrossing-4}{Iterations vs no.~of vertices}]{
\centering
\scalebox{.95}{
\centering
\begin{tikzpicture}
\begin{axis}[axis x line*=bottom,xlabel style={yshift=0.2cm},ylabel style={yshift=-0.2cm},axis y line*=left,legend style={at={(0.425,1.35)},
anchor=north,legend columns=2,draw=none}, width=0.475\textwidth,cycle list name=mylist, mark repeat={5},xtick={10,20,30,40,50,60,70,80,90,100},
xmin=10,
xmax=100,
ytick={0,1000,2000,3000,4000,5000,6000,7000,8000},tick pos=left,
ylabel={Iterations},
xlabel={Number of Vertices},ymajorgrids]
\addplot table [x=avg, y=it rm, col sep=semicolon]{crossingResolutionNorth.csv};
\addplot table [x=avg, y=it rm-1, col sep=semicolon] {crossingResolutionNorth.csv};
\addplot table [x=avg, y=it fa, col sep=semicolon]{crossingResolutionNorth.csv};
\addplot table [x=avg, y=it tr, col sep=semicolon] {crossingResolutionNorth.csv};
\legend{{Unrestricted},{AR-restricted~~},{Huang et al.},{Argyriou et al.}}
\end{axis}
\end{tikzpicture}}}
\caption{Experimental results for the crossing resolution on the AT\&T graphs.}
\label{fig:northCrossing}
\end{figure}

For the crossing resolution, we observed that both variants of our algorithm again outperformed the two force-directed algorithms; see Fig.~\ref{fig:northCrossing-1}. Remarkable is the synchronous behaviour of all four algorithms regarding the crossing resolution, as the curves are nearly parallel. By all these results, we can classify the graphs into ``hard'' or ``easy'' when maximizing their crossing resolution. In particular, graphs with 50 to 70 vertices appear to be harder to improve than graphs with 70 to 80 vertices. Regarding the aspect ratio of the produced drawings, we observe that while our algorithms show a slight increase with the number of vertices, the behaviour for both force-directed algorithms appears to be unstable resulting in a large variance. Again the restricted variant of our algorithm and the two force directed approaches produce drawings with similar aspect ratio, which is much lower than the one of our unrestricted algorithm for larger graphs. All four algorithms behave nearly the same in terms of the number of crossings; see Fig.~\ref{fig:northCrossing-3}. In terms of the number of iterations, we observe that somewhat surprisingly the algorithm of Argyriou et al. converges in the least amount of iterations, while the remaining three algorithms behave nearly the same; see Fig.~\ref{fig:northCrossing-4}.

\begin{figure}[tb]
\centering
\subfloat[\label{fig:northTotal-1}{Total resolution vs no.~of vertices}]{
\centering
\scalebox{.95}{
\centering
\begin{tikzpicture}
\begin{axis}[
axis x line*=bottom,xlabel style={yshift=0.2cm},ylabel style={yshift=-0.2cm},axis y line*=left,
legend style={at={(0.425,1.35)},anchor=north,,legend columns=2,draw=none},
width=0.475\textwidth,
cycle list name=mylist,
mark repeat={5},xlabel={Number of Vertices},ylabel={Total Resolution},
xtick={10,20,30,40,50,60,70,80,90,100},
xmin=10,
xmax=100,
ytick={10,20,30,40,50,60,70,80,90},
tick pos=left,ymajorgrids]
\addplot table [x=avg, y=tr rm, col sep=semicolon]{totalResolutionNorth.csv};
\addplot table [x=avg, y=tr rm-1, col sep=semicolon] {totalResolutionNorth.csv};
\addplot table [x=avg, y=tr fa, col sep=semicolon]{totalResolutionNorth.csv};
\addplot table [x=avg, y=tr tr, col sep=semicolon] {totalResolutionNorth.csv};
\legend{{Unrestricted},{AR-restricted~~},{Huang et al.},{Argyriou et al.}}
\end{axis}
\end{tikzpicture}}}
\subfloat[\label{fig:northTotal-2}{Aspect ratio vs no.~of vertices}]{
\centering
\scalebox{.95}{
\centering
\begin{tikzpicture}
\begin{axis}[axis x line*=bottom,xlabel style={yshift=0.2cm},axis y line*=left,legend style={at={(0.425,1.35)},
anchor=north,legend columns=2,draw=none},ylabel style={yshift=-0.2cm}, width=0.475\textwidth,cycle list name=mylist, mark repeat={5},xtick={0,10,20,30,40,50,60,70,80,90,100},ylabel={Aspect Ratio},
xlabel={Number of Vertices},
xmin=10,
xmax=100,
ytick={1,2,5,10,20,50,100,200},ymode=log, log ticks with fixed point,tick pos=left, ymajorgrids,]
\addplot table [x=avg, y=ar rm, col sep=semicolon]{totalResolutionNorth.csv};
\addplot table [x=avg, y=ar rm-1, col sep=semicolon] {totalResolutionNorth.csv};
\addplot table [x=avg, y=ar fa, col sep=semicolon]{totalResolutionNorth.csv};
\addplot table [x=avg, y=ar tr, col sep=semicolon] {totalResolutionNorth.csv};
\legend{{Unrestricted},{AR-restricted~~},{Huang et al.},{Argyriou et al.}}
\end{axis}
\end{tikzpicture}}}

\subfloat[\label{fig:northTotal-3}{No.~of crossings vs no.~of vertices}]{
\centering
\scalebox{.95}{
\centering
\begin{tikzpicture}
\begin{axis}[axis x line*=bottom,xlabel style={yshift=0.2cm},axis y line*=left,legend style={at={(0.425,1.35)},
anchor=north,legend columns=2,draw=none}, width=0.475\textwidth,cycle list name=mylist,  mark repeat={5},xtick={10,20,30,40,50,60,70,80,90,100},
xmin=10,
xmax=100,
ytick={0,250,500,750,1000,1250,1500,1750},ylabel={Number of Crossings},
xlabel={Number of Vertices},tick pos=left,ylabel style={yshift=-0.2cm}, ymajorgrids]
\addplot table [x=avg, y=cn rm, col sep=semicolon]{totalResolutionNorth.csv};
\addplot table [x=avg, y=cn rm-1, col sep=semicolon] {totalResolutionNorth.csv};
\addplot table [x=avg, y=cn fa, col sep=semicolon]{totalResolutionNorth.csv};
\addplot table [x=avg, y=cn tr, col sep=semicolon] {totalResolutionNorth.csv};
\legend{{Unrestricted},{AR-restricted~~},{Huang et al.},{Argyriou et al.}}
\end{axis}
\end{tikzpicture}}}
\subfloat[\label{fig:northTotal-4}{Iterations vs no.~of vertices}]{
\centering
\scalebox{.95}{
\centering
\begin{tikzpicture}
\begin{axis}[axis x line*=bottom,xlabel style={yshift=0.2cm},ylabel style={yshift=-0.2cm},axis y line*=left,legend style={at={(0.425,1.35)},
anchor=north,legend columns=2,draw=none}, width=0.475\textwidth,cycle list name=mylist, mark repeat={5},xtick={10,20,30,40,50,60,70,80,90,100},
xmin=10,
xmax=100,
ytick={0,1000,2000,3000,4000,5000,6000,7000,8000},tick pos=left,
ylabel={Iterations},
xlabel={Number of Vertices},ymajorgrids]
\addplot table [x=avg, y=it rm, col sep=semicolon]{totalResolutionNorth.csv};
\addplot table [x=avg, y=it rm-1, col sep=semicolon] {totalResolutionNorth.csv};
\addplot table [x=avg, y=it fa, col sep=semicolon]{totalResolutionNorth.csv};
\addplot table [x=avg, y=it tr, col sep=semicolon] {totalResolutionNorth.csv};
\legend{{Unrestricted},{AR-restricted~~},{Huang et al.},{Argyriou et al.}}
\end{axis}
\end{tikzpicture}}}
\caption{Experimental results for the total resolution on the AT\&T graphs.}
\label{fig:northTotal}
\end{figure}
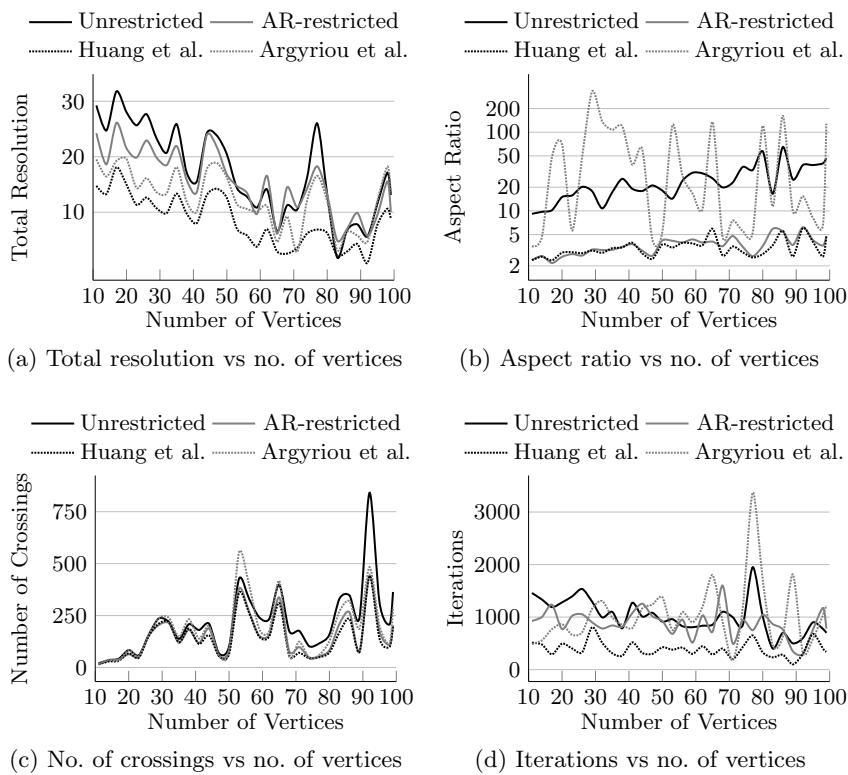

In the total resolution experiment, we observed similar results as in the experiment on the Rome graphs for small graphs, that is, our unrestricted algorithm outperforms the other three ones, while the restricted variant of our algorithm yields drawings and then the algorithm by Argyriou et al.; see Fig.~\ref{fig:northTotal-1}. For larger graphs, however, these three algorithms achieve similar results while still outperforming the algorithm by Huang et al. The results for the aspect ratio and number of crossings are similar to those of the crossing resolution experiment, with the exception of the fact that the algorithm of Huang et al. performs more stable with respect to the aspect ratio; see Figs.~\ref{fig:northTotal-2} and~\ref{fig:northTotal-3}. With respect to the number of iterations, our two algorithms and the one by Argyriou et al. show similar behavior needing more iterations than the algorithm by Huang et al.\ in order to converge; see Fig.~\ref{fig:northTotal-4}. 

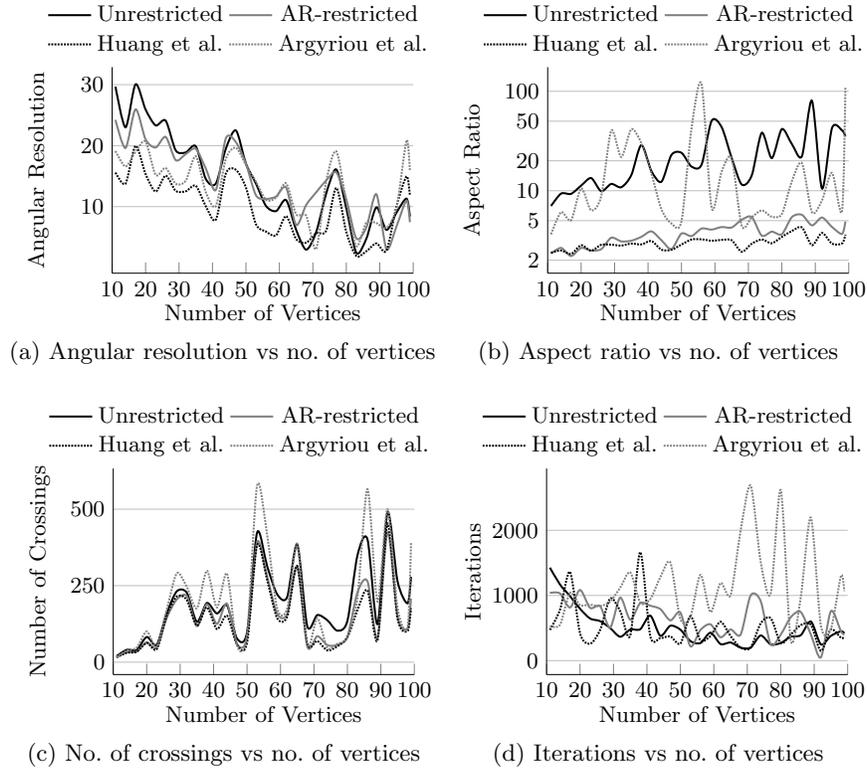
\begin{figure}[htb]
\centering
\subfloat[\label{fig:northAngular-1}{Angular resolution vs no.~of vertices}]{
\centering
\scalebox{.95}{
\centering
\begin{tikzpicture}
\begin{axis}[
axis x line*=bottom,xlabel style={yshift=0.2cm},ylabel style={yshift=-0.2cm},axis y line*=left,
legend style={at={(0.425,1.35)},anchor=north,,legend columns=2,draw=none},
width=0.475\textwidth,
cycle list name=mylist,
mark repeat={5},xlabel={Number of Vertices},ylabel={Angular Resolution},
xtick={10,20,30,40,50,60,70,80,90,100},
xmin=10,
xmax=100,
ytick={10,20,30,40,50,60,70,80,90},
tick pos=left,ymajorgrids]
\addplot table [x=avg, y=an rm, col sep=semicolon]{angularResolutionNorth.csv};
\addplot table [x=avg, y=an rm-1, col sep=semicolon] {angularResolutionNorth.csv};
\addplot table [x=avg, y=an fa, col sep=semicolon]{angularResolutionNorth.csv};
\addplot table [x=avg, y=an tr, col sep=semicolon] {angularResolutionNorth.csv};
\legend{{Unrestricted},{AR-restricted~~},{Huang et al.},{Argyriou et al.}}
\end{axis}
\end{tikzpicture}}}
\subfloat[\label{fig:northAngular-2}{Aspect ratio vs no.~of vertices}]{
\centering
\scalebox{.95}{
\centering
\begin{tikzpicture}
\begin{axis}[axis x line*=bottom,xlabel style={yshift=0.2cm},axis y line*=left,legend style={at={(0.425,1.35)},
anchor=north,legend columns=2,draw=none},ylabel style={yshift=-0.2cm}, width=0.475\textwidth,cycle list name=mylist, mark repeat={5},xtick={0,10,20,30,40,50,60,70,80,90,100},ylabel={Aspect Ratio},
xlabel={Number of Vertices},
xmin=10,
xmax=100,
ytick={1,2,5,10,20,50,100,200},ymode=log, log ticks with fixed point,tick pos=left, ymajorgrids,]
\addplot table [x=avg, y=ar rm, col sep=semicolon]{angularResolutionNorth.csv};
\addplot table [x=avg, y=ar rm-1, col sep=semicolon] {angularResolutionNorth.csv};
\addplot table [x=avg, y=ar fa, col sep=semicolon]{angularResolutionNorth.csv};
\addplot table [x=avg, y=ar tr, col sep=semicolon] {angularResolutionNorth.csv};
\legend{{Unrestricted},{AR-restricted~~},{Huang et al.},{Argyriou et al.}}
\end{axis}
\end{tikzpicture}}}

\subfloat[\label{fig:northAngular-3}{No.~of crossings vs no.~of vertices}]{
\centering
\scalebox{.95}{
\centering
\begin{tikzpicture}
\begin{axis}[axis x line*=bottom,xlabel style={yshift=0.2cm},axis y line*=left,legend style={at={(0.425,1.35)},
anchor=north,legend columns=2,draw=none}, width=0.475\textwidth,cycle list name=mylist,  mark repeat={5},xtick={10,20,30,40,50,60,70,80,90,100},
xmin=10,
xmax=100,
ytick={0,250,500,750,1000,1250,1500,1750},ylabel={Number of Crossings},
xlabel={Number of Vertices},tick pos=left,ylabel style={yshift=-0.2cm}, ymajorgrids]
\addplot table [x=avg, y=cn rm, col sep=semicolon]{angularResolutionNorth.csv};
\addplot table [x=avg, y=cn rm-1, col sep=semicolon] {angularResolutionNorth.csv};
\addplot table [x=avg, y=cn fa, col sep=semicolon]{angularResolutionNorth.csv};
\addplot table [x=avg, y=cn tr, col sep=semicolon] {angularResolutionNorth.csv};
\legend{{Unrestricted},{AR-restricted~~},{Huang et al.},{Argyriou et al.}}
\end{axis}
\end{tikzpicture}}}
\subfloat[\label{fig:northAngular-4}{Iterations vs no.~of vertices}]{
\centering
\scalebox{.95}{
\centering
\begin{tikzpicture}
\begin{axis}[axis x line*=bottom,xlabel style={yshift=0.2cm},ylabel style={yshift=-0.2cm},axis y line*=left,legend style={at={(0.425,1.35)},
anchor=north,legend columns=2,draw=none}, width=0.475\textwidth,cycle list name=mylist, mark repeat={5},xtick={10,20,30,40,50,60,70,80,90,100},
xmin=10,
xmax=100,
ytick={0,1000,2000,3000,4000,5000,6000,7000,8000},tick pos=left,
ylabel={Iterations},
xlabel={Number of Vertices},ymajorgrids]
\addplot table [x=avg, y=it rm, col sep=semicolon]{angularResolutionNorth.csv};
\addplot table [x=avg, y=it rm-1, col sep=semicolon] {angularResolutionNorth.csv};
\addplot table [x=avg, y=it fa, col sep=semicolon]{angularResolutionNorth.csv};
\addplot table [x=avg, y=it tr, col sep=semicolon] {angularResolutionNorth.csv};
\legend{{Unrestricted},{AR-restricted~~},{Huang et al.},{Argyriou et al.}}
\end{axis}
\end{tikzpicture}}}
\caption{Experimental results for the angular resolution on the AT\&T graphs.}
\label{fig:northAngular}
\end{figure}

In the angular resolution experiment, we again obtain a not-so-clear picture concerning the ranking of the algorithms, especially for higher number of vertices the ranking varies; see Fig.~\ref{fig:northAngular-1}. Only the algorithm by Huang et al. seems to be mostly at the last rank. Concerning the aspect ratio we see very good behaviour for our restricted variant and the algorithm by Huang et al. while the remaining two algorithms show large variance and much worse values; see Fig.~\ref{fig:northAngular-2}. For the number of crossings, we again observe that all algorithms achieve similar values, however, our unrestricted algorithm and the algorithm by Argyriou et al. achieve slightly higher values for larger graphs; see Fig.~\ref{fig:northAngular-3}. Finally, both our algorithms and the one by Huang et al. need a similar number of iterations for convergence which is lower than the one by Argyriou et al.; see Fig.~\ref{fig:northAngular-4}.

Summarizing, we conclude that compared to the Rome graphs, the AT\&T graphs show a much higher variance regarding the various resolution measures. 

\begin{table}[t!]

\caption{Summary of the results for the Graph Drawing Contest Graphs.}
\centering
\begin{tabular}{c|c|c|c|c}
\toprule
~~Graph~~ & ~~CoffeeVM~~ & ~~TuebingenMidnight~~ & ~~Time restricted~~ & ~~Our best~~ \\
\midrule
1 & 90  & 77  & 89.99 & 89.99\\
2 & 88.23  & 42 & 88.21 & 88.7 \\
3 & 90  & 89 & 87.86 & 89.95 \\
4 & 88.97  & 89 & 77.13  & 89.05 \\
5 & 80.4  &  30 & 78.68 & 86.96 \\
6 & 90  & 78 & 89.96 & 89.96\\
7 & 56.537  & 34 & 55.77  & 63.62 \\
8 & 84.95  & 61 & 81.18 & 89.28 \\
9 & 59.885  & 9 & 54.63 & 88.2 \\
10 & 20.978  & 4 & 23.60  & 23.72 \\
11 & 46.684  & 6 & 57.00 & 72.00 \\
12 & 36.47  & 5 & 26.24 & 35.86 \\
13 & 25.456  & 4 & 22.43 & 33.68 \\
14 & 33.52  & 5 & 29.69 & 43.08 \\
15 & 20.512  & 4 & 13.51 & 29.18 \\
\bottomrule
\end{tabular}
\label{tab:gdContest2017}
\end{table}

\section{Graph Drawing Contest 2017 Graphs}
\label{app:contest}

Our primary motivation for this work was our participation to the Graph Drawing contest in 2017, where we miserably performed\footnote{\texttt{http://www.graphdrawing.de/contest2017/results.html}}; the topic was the maximization of the crossing resolution. Our approach for the contest in 2017 was a mixture of the two force directed algorithms by Argyriou et al. and Huang et al.

We give a comparison of our new approach to the performances of the clear winner ``CoffeeVM'' of last year's graph drawing contest and our previous team ``TuebingenMidnight'' in Table~\ref{tab:gdContest2017}. Note that in the contest the teams had only one hour to compute layouts for all 15 contest graphs. For our algorithm, we provide results that were achieved with the same time limit (see column ``Time restricted''), as well as our best results which were achieved without a strict time limit (see column ``Our best'').

We can observe that for almost all graphs, our new approach achieves only slightly worse results than the ones of the last year's contest winner. On a few graphs (namely, graphs 10 and 11), we even achieve better results. With a single exception (namely, graph 4), we easily outperformed our results from last year.
If we neglect the time restriction, for all the graphs, the results  are (sometimes considerably) better than or at least about the same as last year's contest winner. We can conclude that our new approach has good potential for the application in this year's graph drawing contest, however, we also note that more careful graph dependent parameterization will be needed to compute competitive solutions within the provided time.


\section{Sample Drawings}
\label{app:samples}

In this section, we present drawings of the 5th and of the 9th graph given in the Graph Drawing contest 2017 that are produced by different variants of our algorithm and of the algorithms by Argyriou et al.~\cite{DBLP:journals/cj/ArgyriouBS13}, and by Huang et al.~\cite{DBLP:journals/vlc/HuangEHL13}; see Figures~\ref{fig:graph5} and~\ref{fig:graph9}, respectively. Each variant was obtained by adjusting each of the aforementioned algorithms to optimize the crossing resolution, the angular resolution and the total resolution, respectively; the aesthetic criterion optimized by each variant is reported in the caption of its corresponding subfigure. As initial drawings for all algorithms, we used the layouts shown in Figure~\ref{fig:graph5And9-yfiles} that we computed with the SmartOrganic layouter of the yFiles library~\cite{DBLP:books/sp/04/WieseE004}.

\begin{figure}[htbp]
\centering
\subfloat[\label{fig:graph5-yfiles}{}]{
\centering
\includegraphics[width=0.3\textwidth]{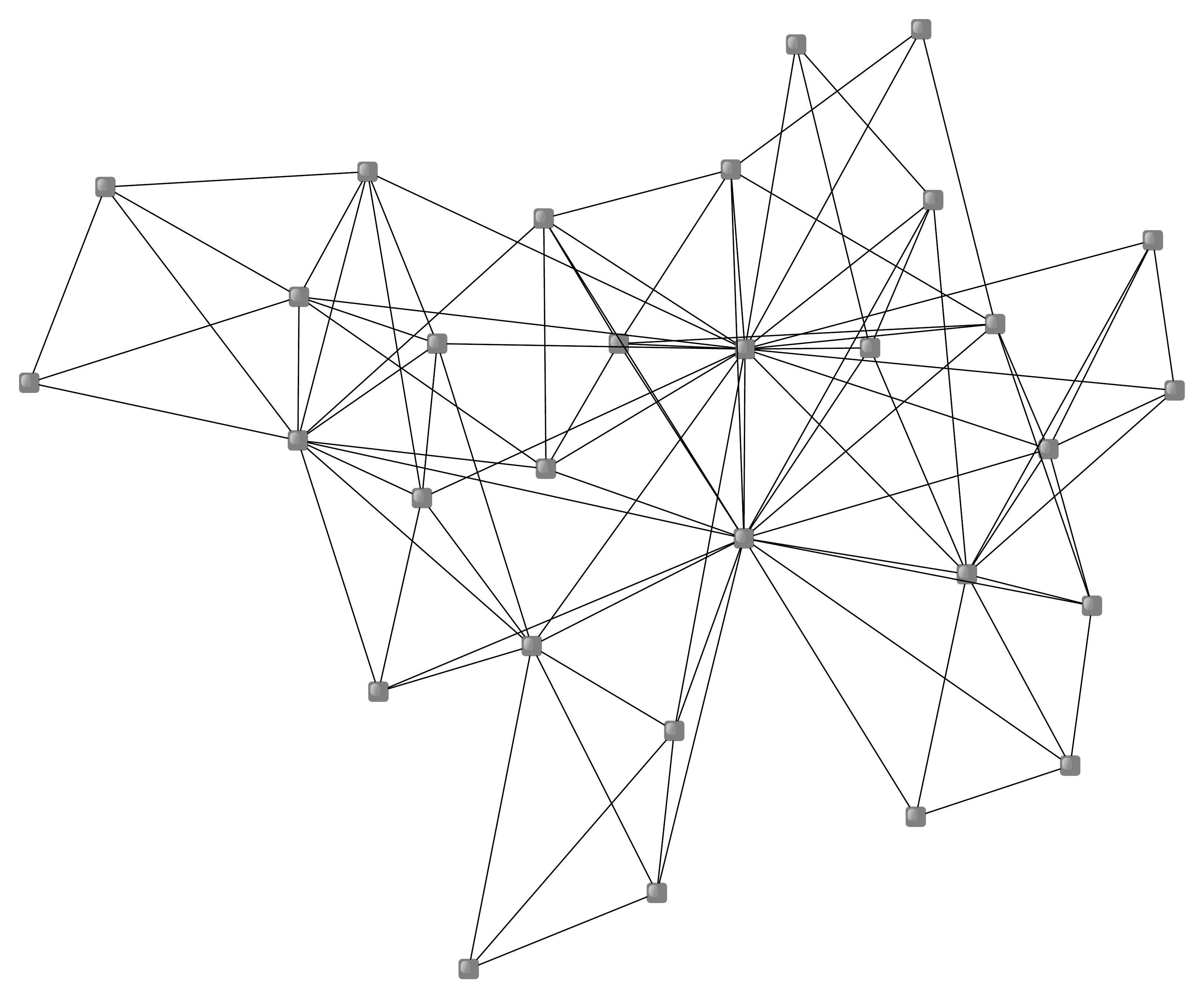}
}
\hfil
\subfloat[\label{fig:graph9-yfiles}{}]{
\centering
\includegraphics[width=0.3\textwidth]{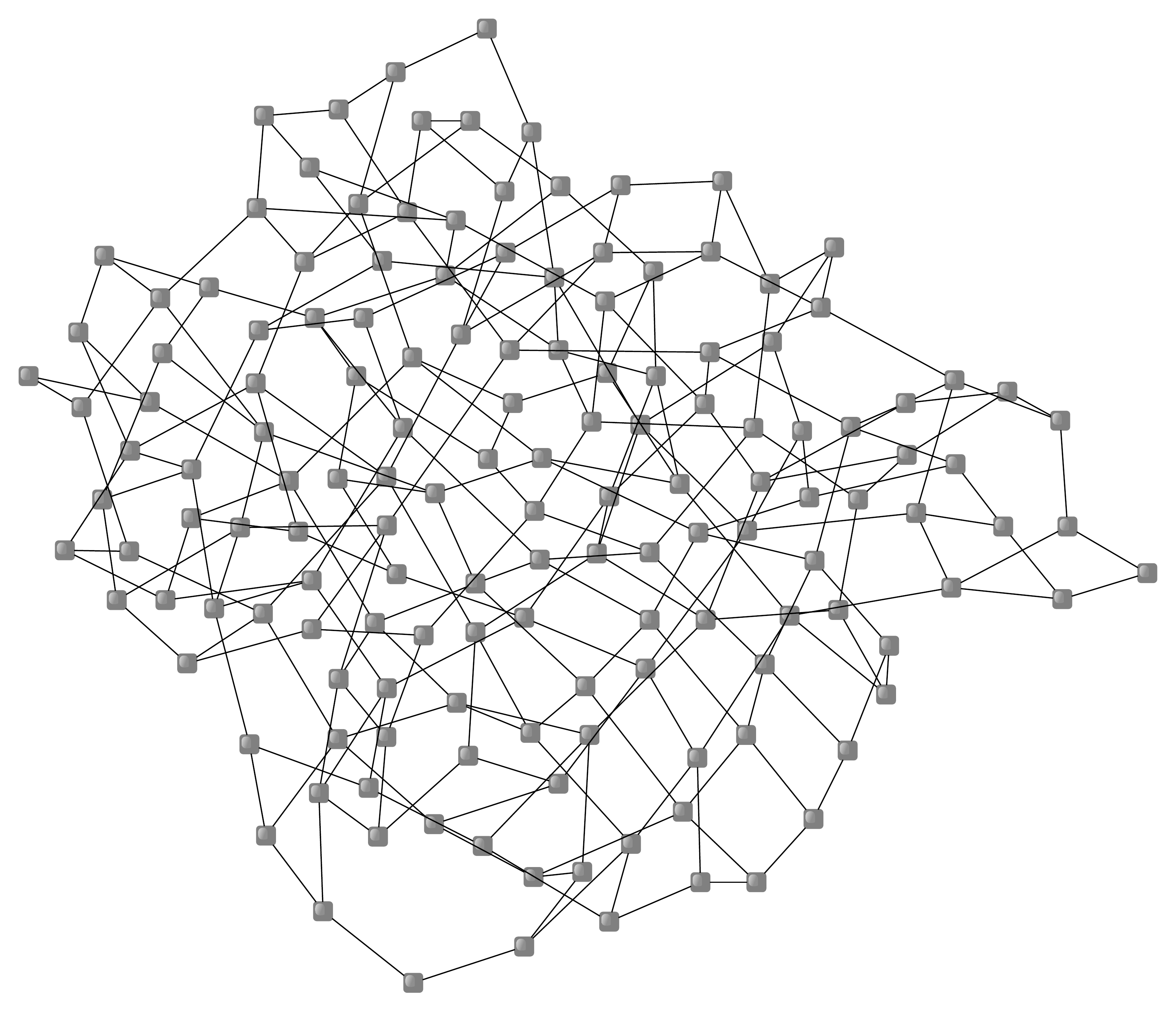}
}
\caption{Illustration of 
(a)~the 5th, and of
(b)~the 9th graph of the Graph Drawing contest 2017.}
\label{fig:graph5And9-yfiles}
\end{figure}

\begin{figure}[p]
\centering
\subfloat[\label{fig:graph5-ourBest-cr}{Crossing Resolution}]{
\centering
\includegraphics[width=0.3\textwidth]{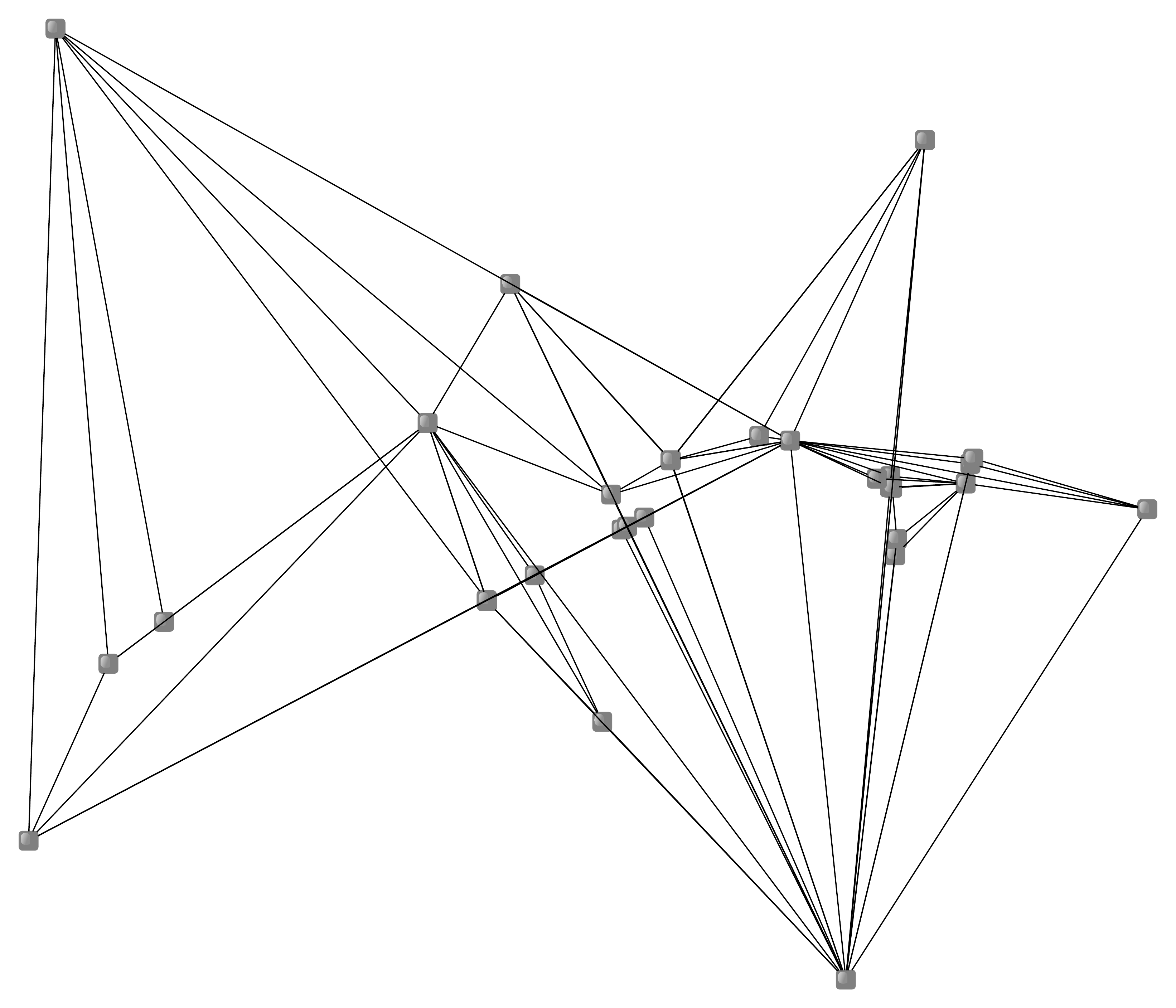}
}
\hfill
\subfloat[\label{fig:graph5-ourBest-ar}{Angular Resolution}]{
\centering
\includegraphics[width=0.3\textwidth]{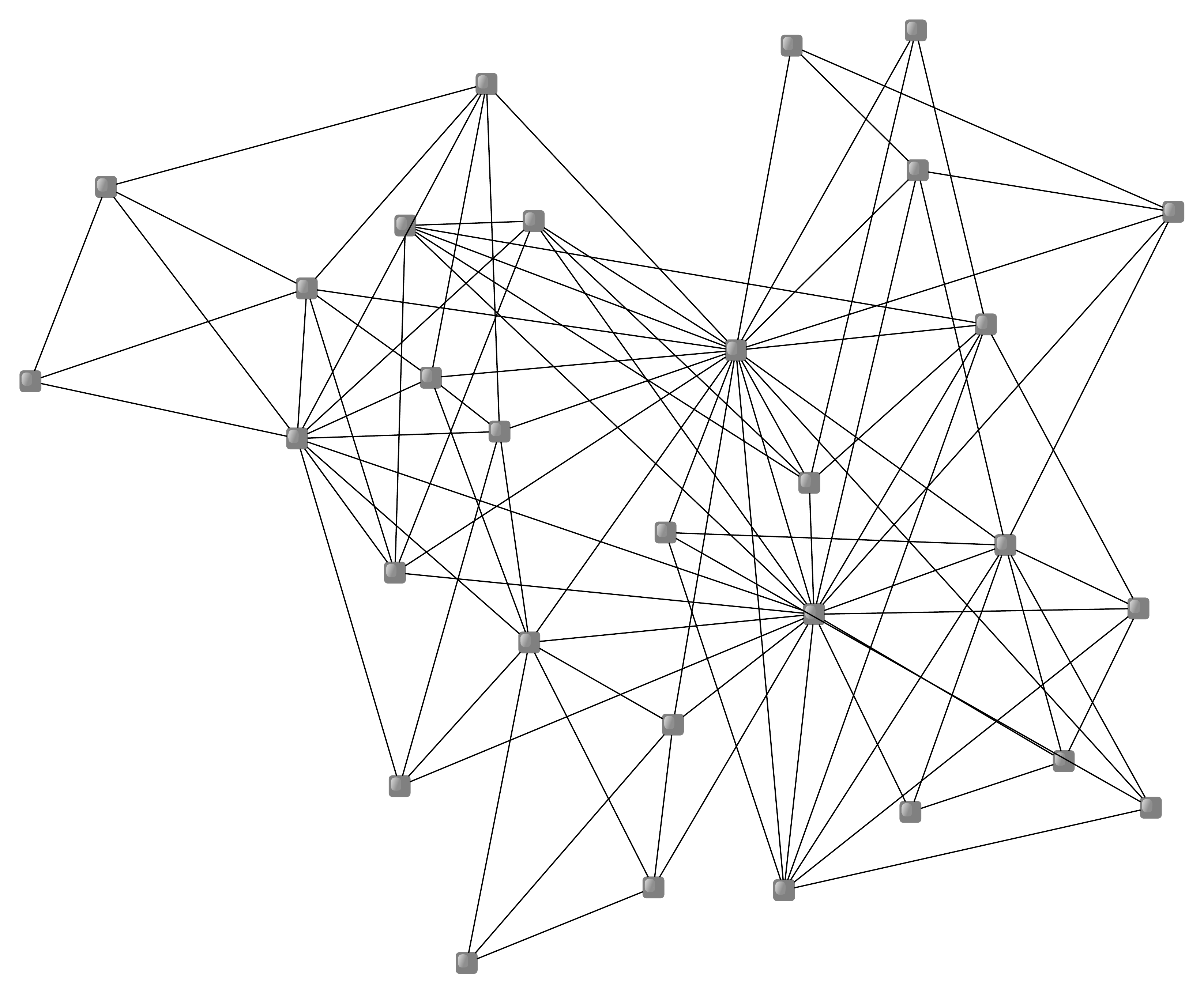}
}
\hfill
\subfloat[\label{fig:graph5-ourBest-tr}{Total Resolution}]{
\centering
\includegraphics[width=0.3\textwidth]{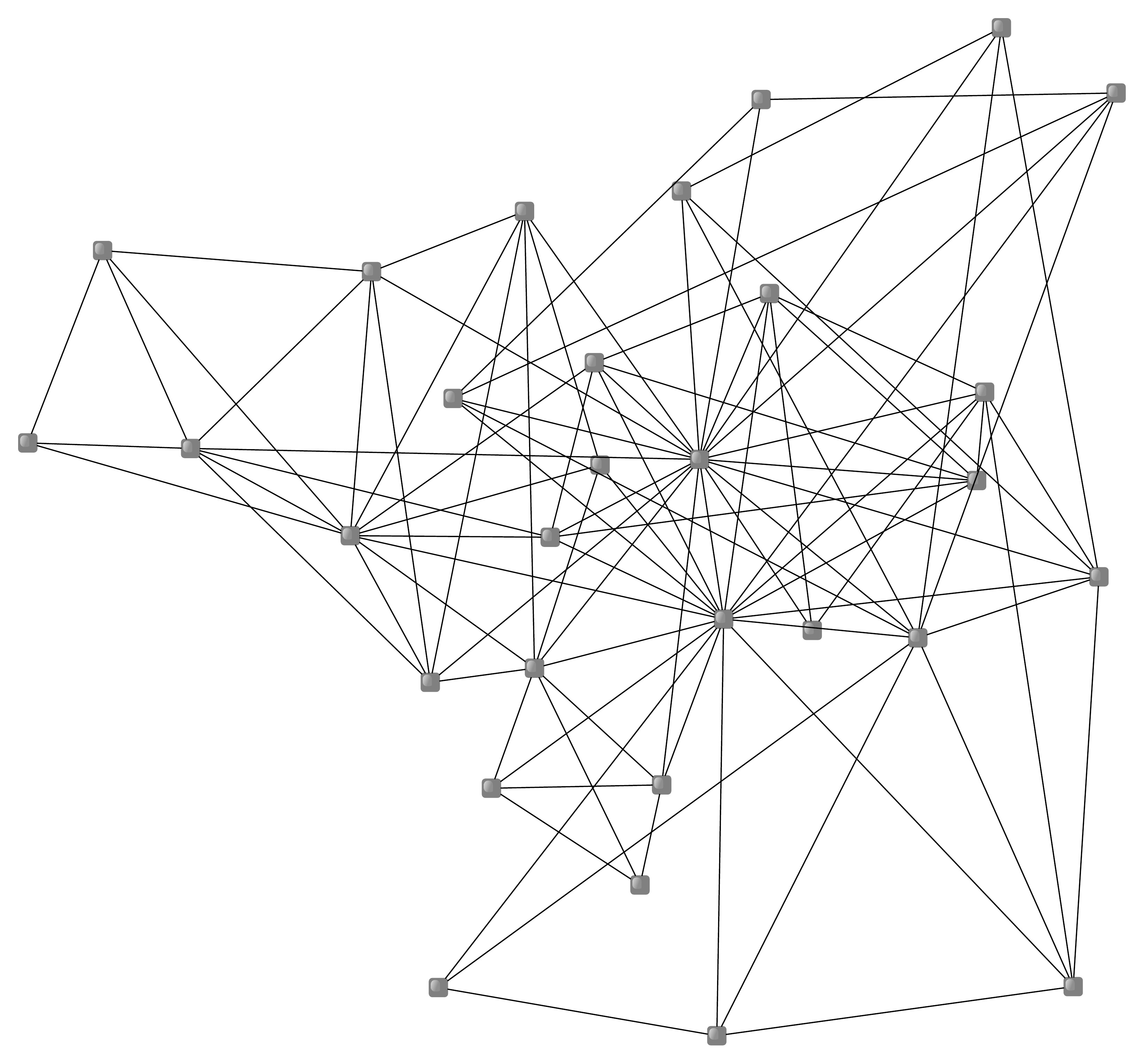}
}
\hfill
\subfloat[\label{fig:graph5-ourRestricted-cr}{Crossing Resolution}]{
\centering
\includegraphics[width=0.3\textwidth]{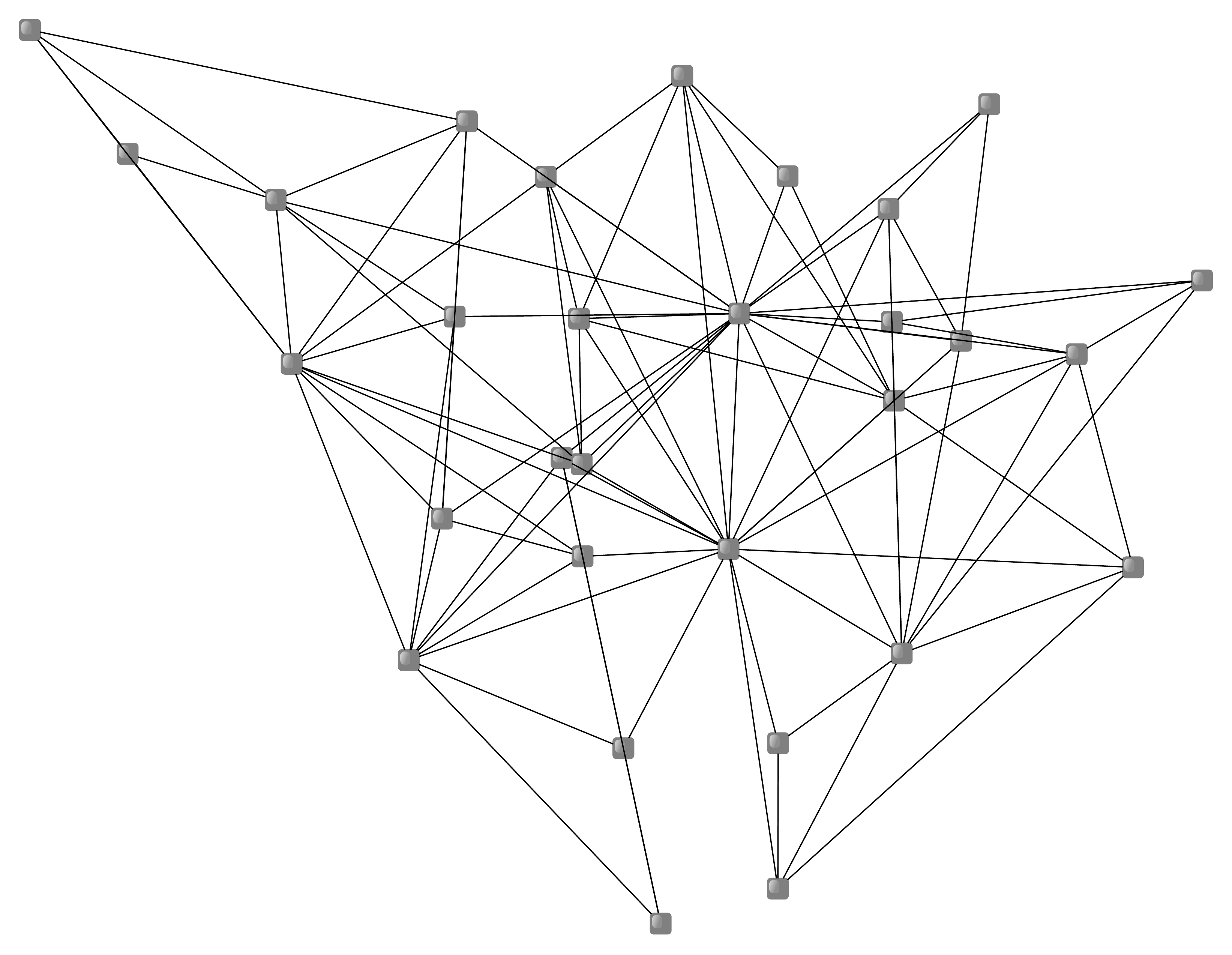}
}
\hfill
\subfloat[\label{fig:graph5-ourRestricted-ar}{Angular Resolution}]{
\centering
\includegraphics[width=0.3\textwidth]{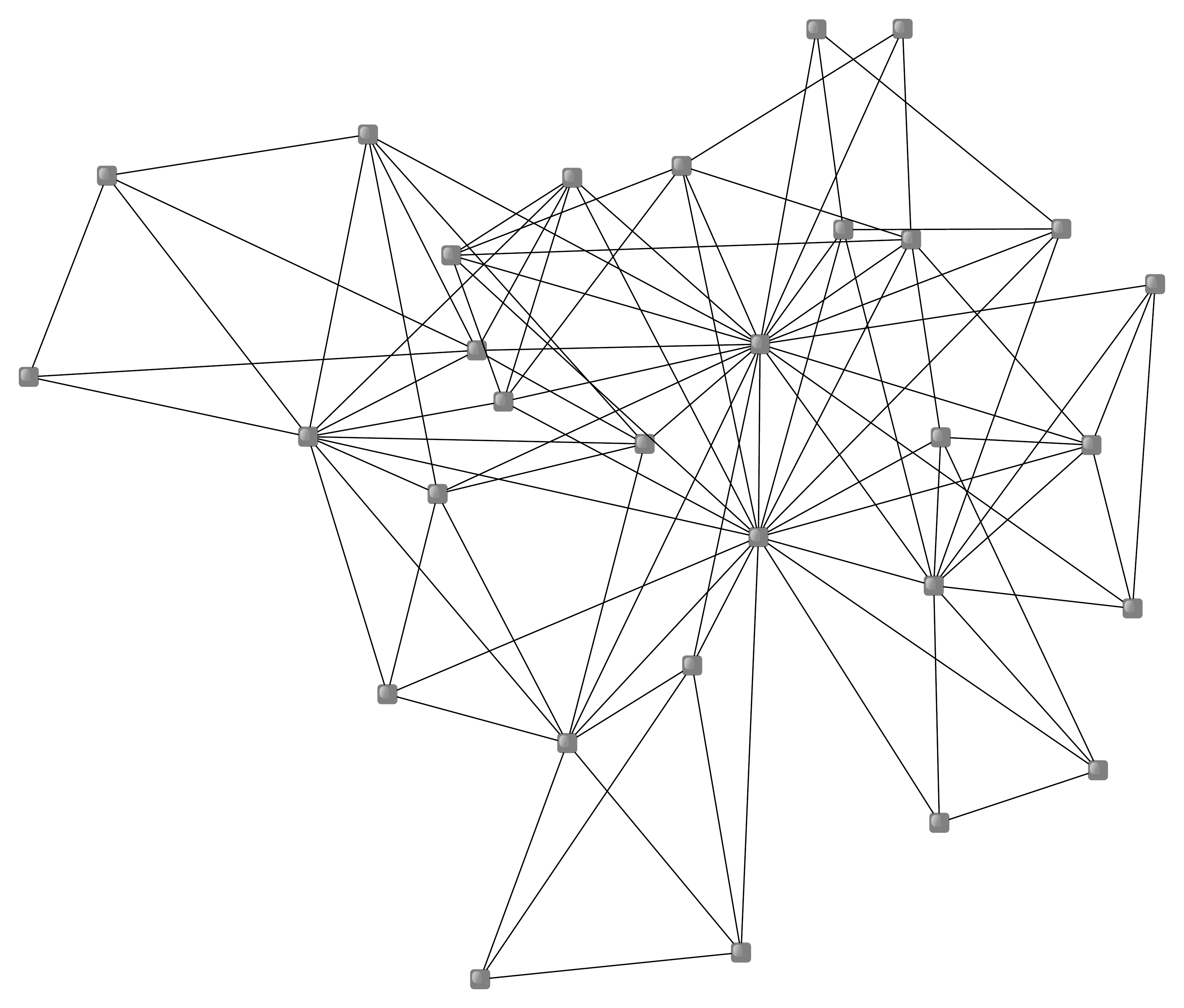}
}
\hfill
\subfloat[\label{fig:graph5-ourRestricted-tr}{Total Resolution}]{
\centering
\includegraphics[width=0.3\textwidth]{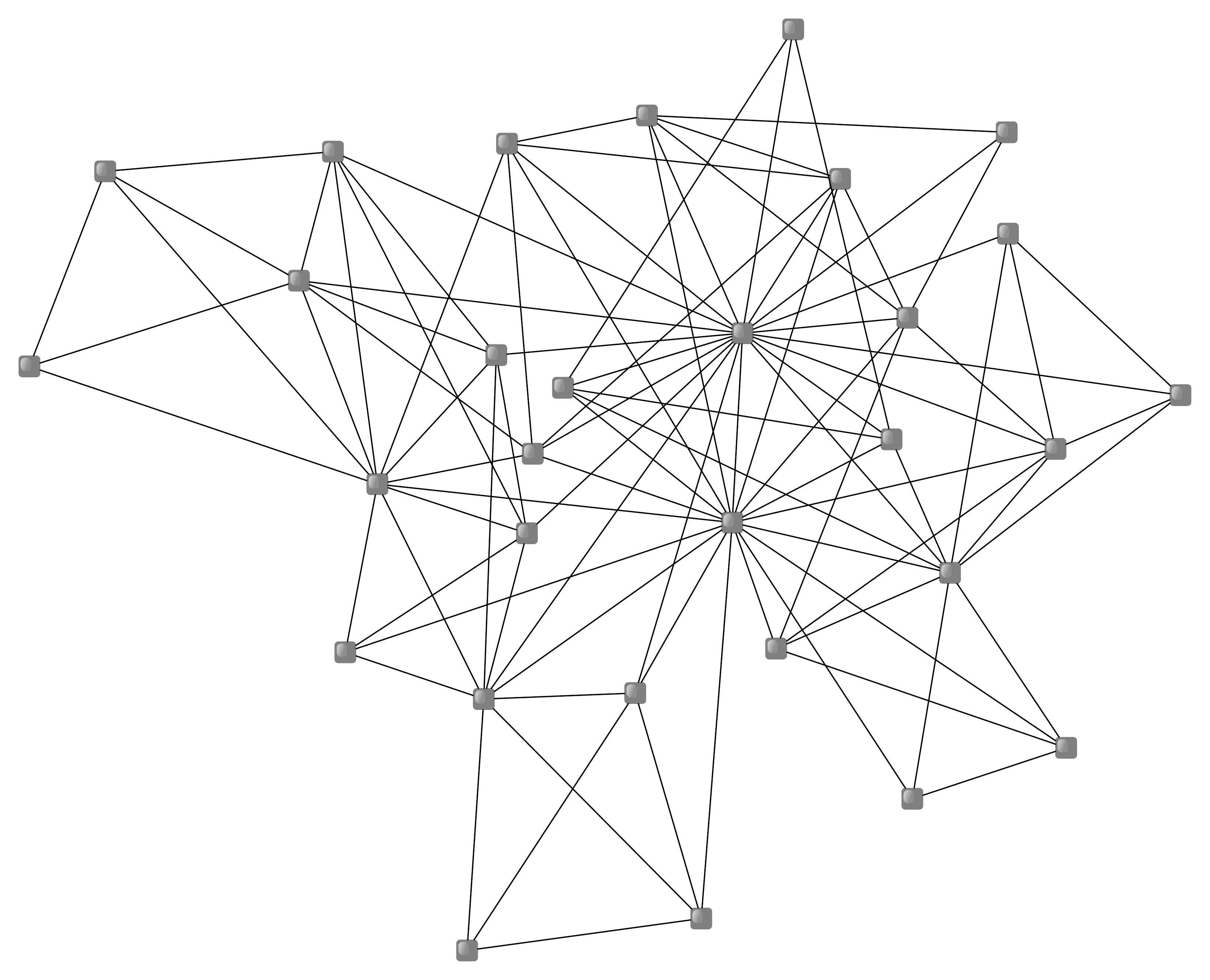}
}
\hfill
\subfloat[\label{fig:graph5-argyriou-cr}{Crossing Resolution}]{
\centering
\includegraphics[width=0.3\textwidth]{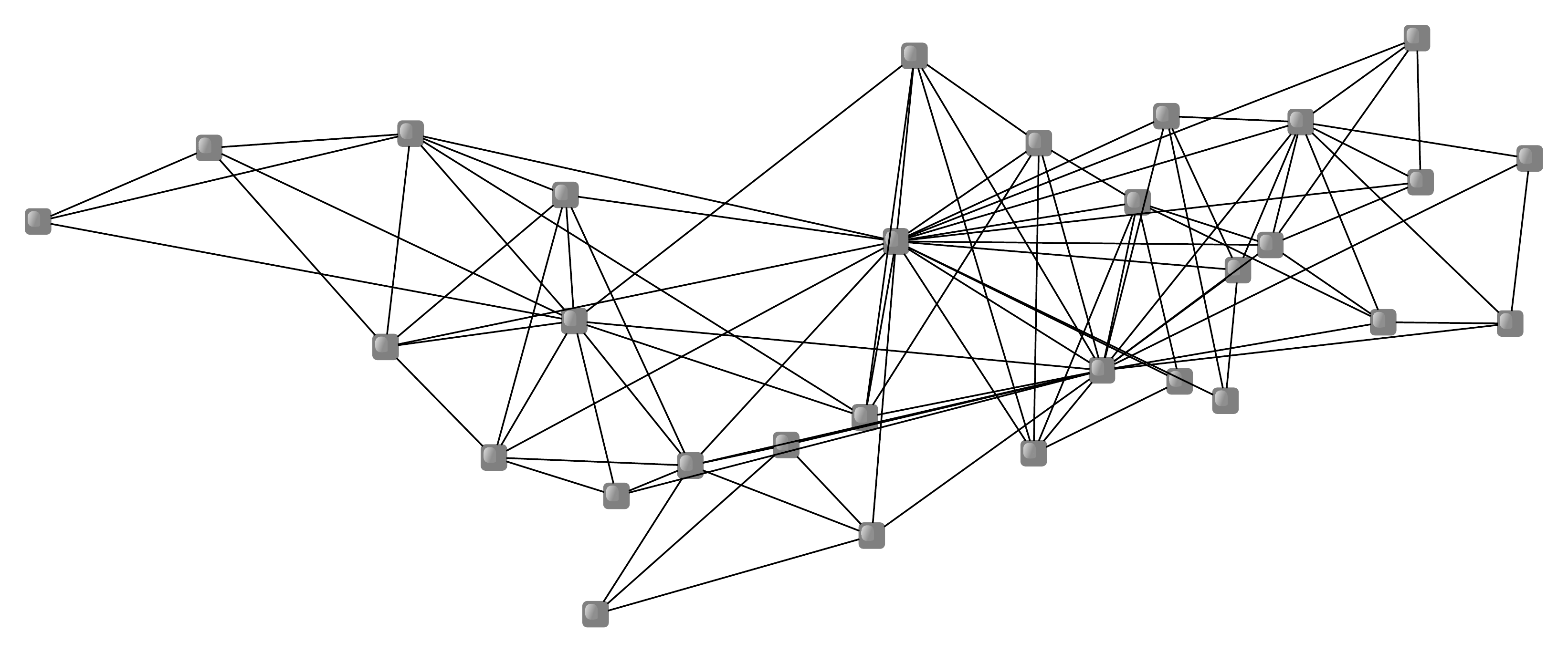}
}
\hfill
\subfloat[\label{fig:graph5-argyriou-ar}{Angular Resolution}]{
\centering
\includegraphics[width=0.3\textwidth]{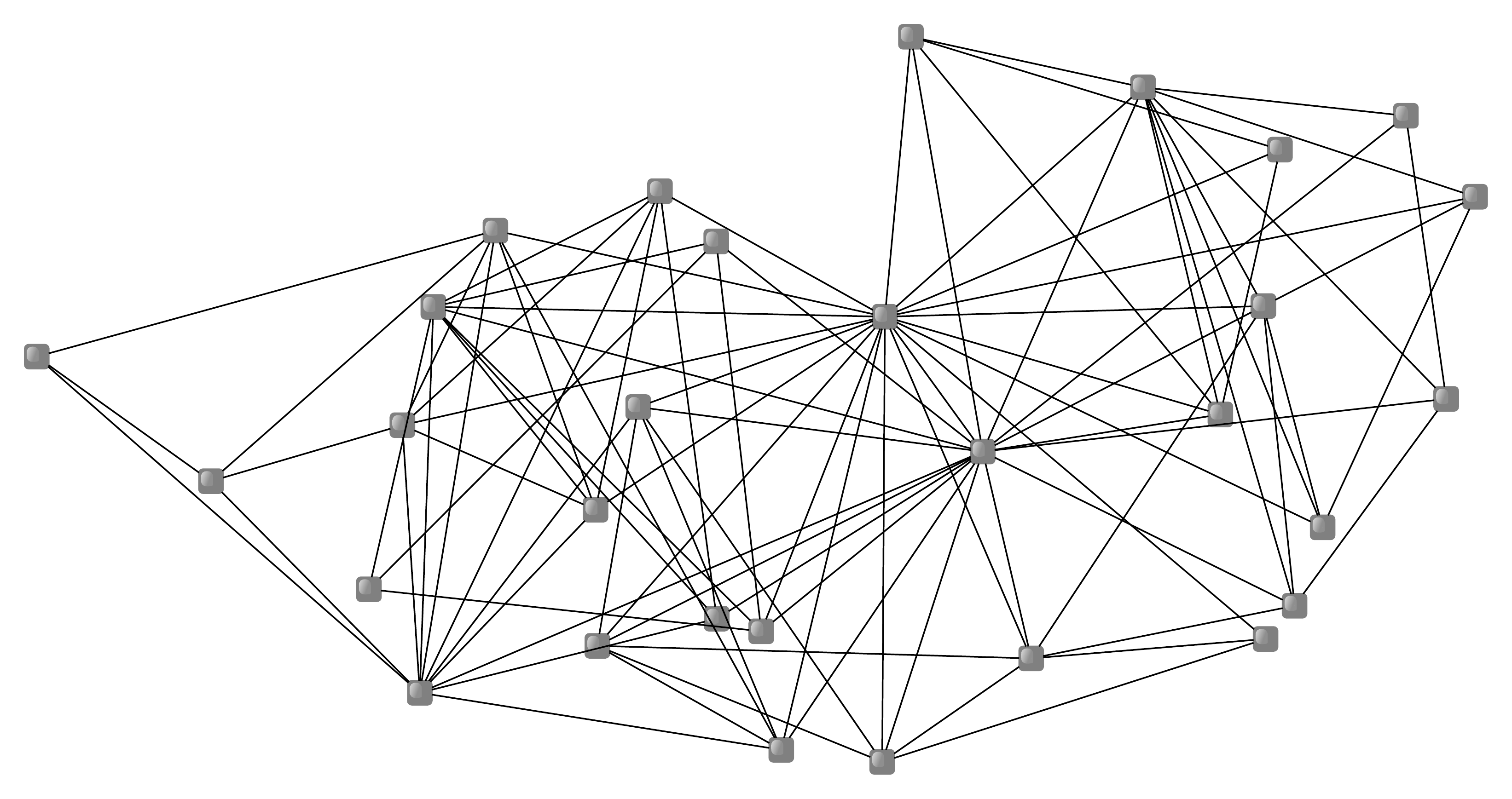}
}
\hfill
\subfloat[\label{fig:graph5-argyriou-tr}{Total Resolution}]{
\centering
\includegraphics[width=0.3\textwidth]{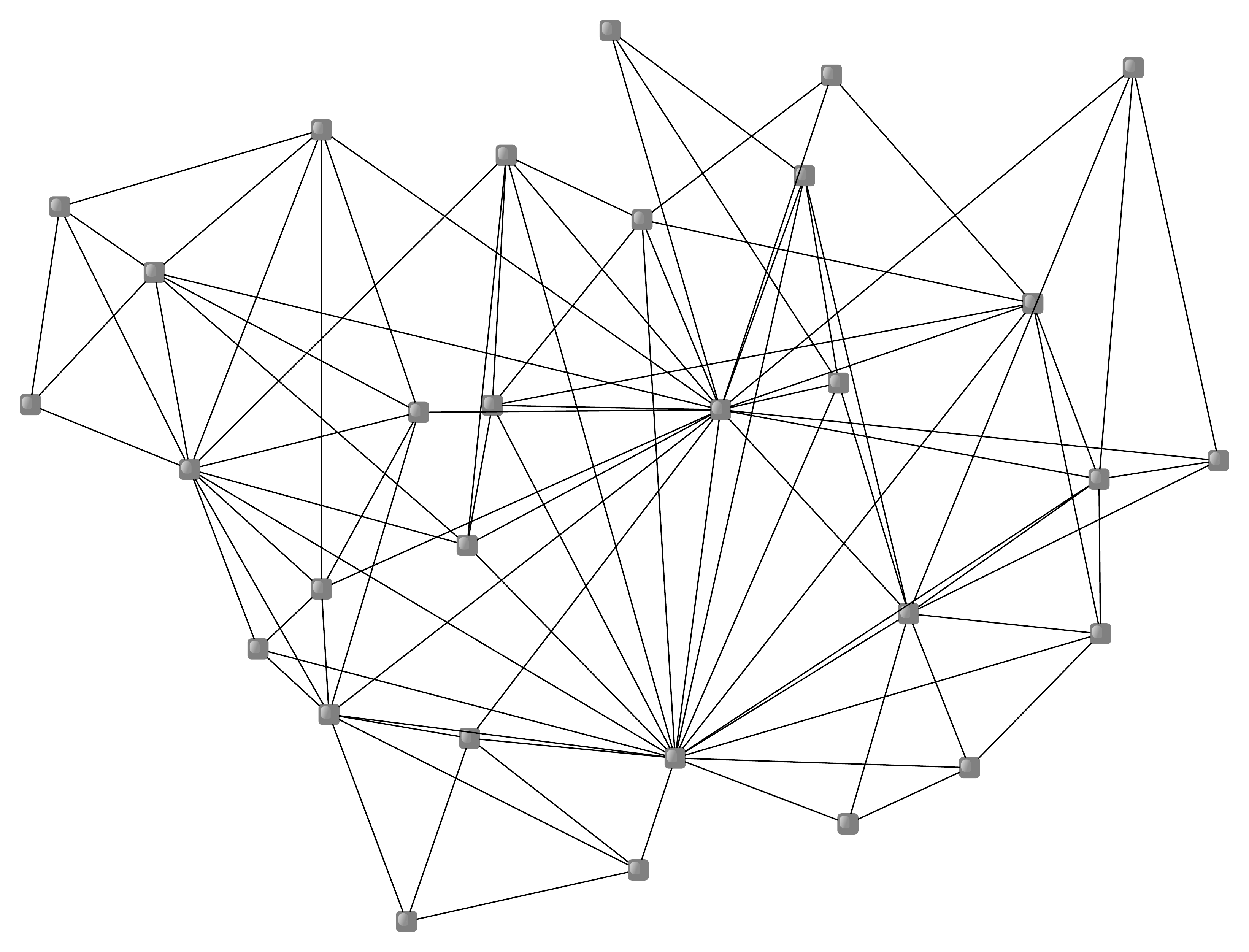}
}
\hfill
\subfloat[\label{fig:graph5-huang-cr}{Crossing Resolution}]{
\centering
\includegraphics[width=0.3\textwidth]{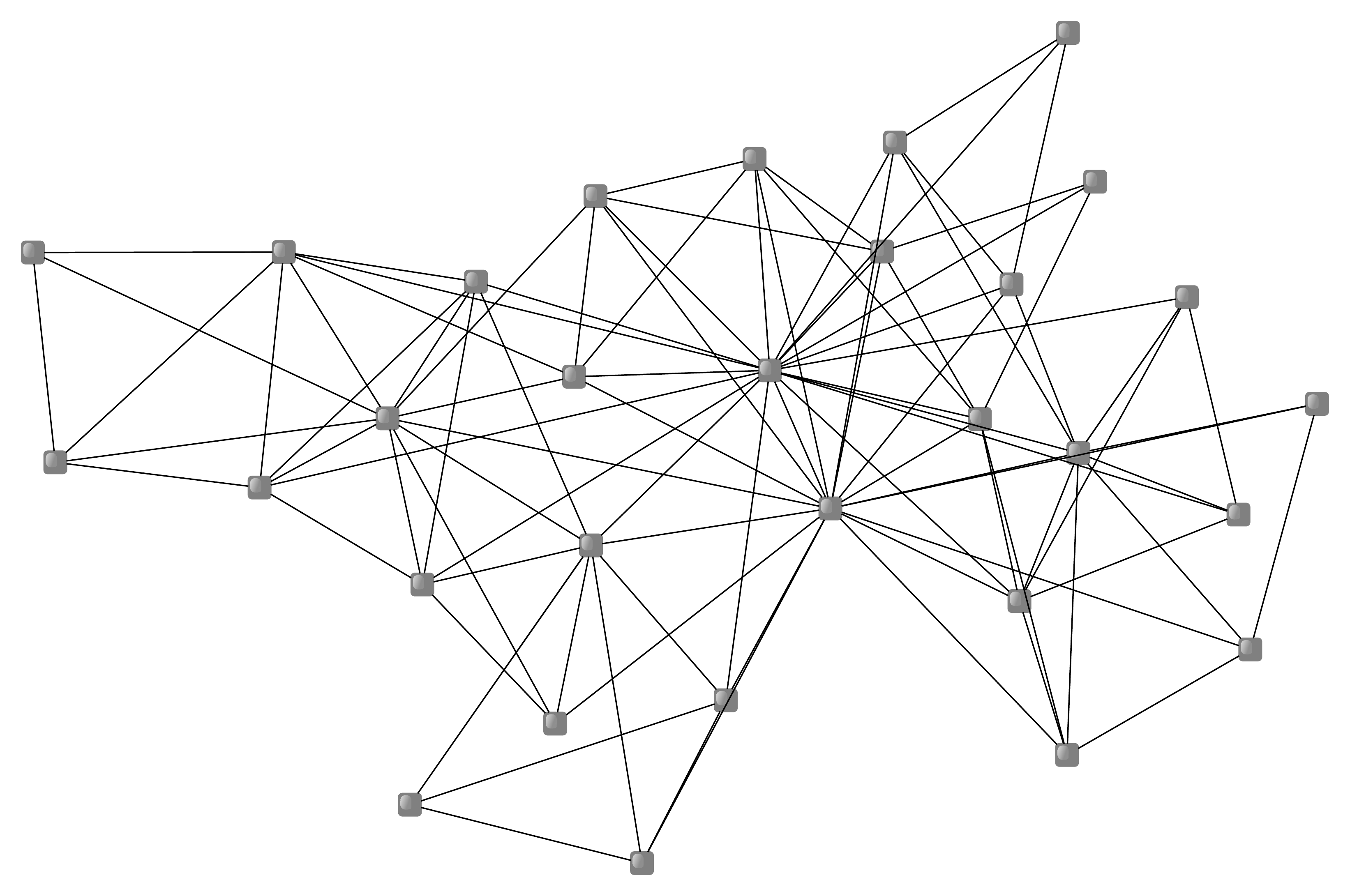}
}
\hfill
\subfloat[\label{fig:graph5-huang-ar}{Angular Resolution}]{
\centering
\includegraphics[width=0.3\textwidth]{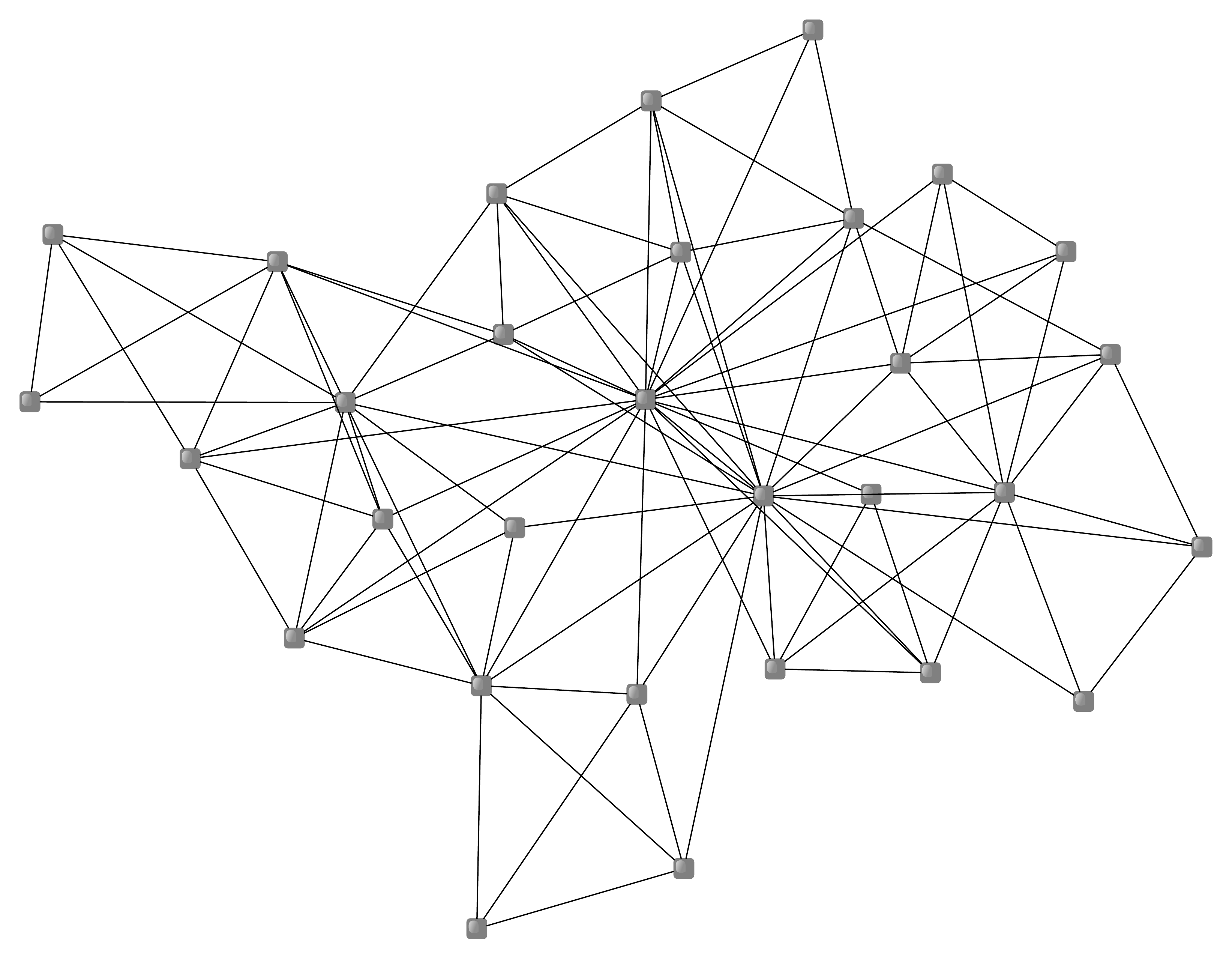}
}
\hfill
\subfloat[\label{fig:graph5-huang-tr}{Total Resolution}]{
\centering
\includegraphics[width=0.3\textwidth]{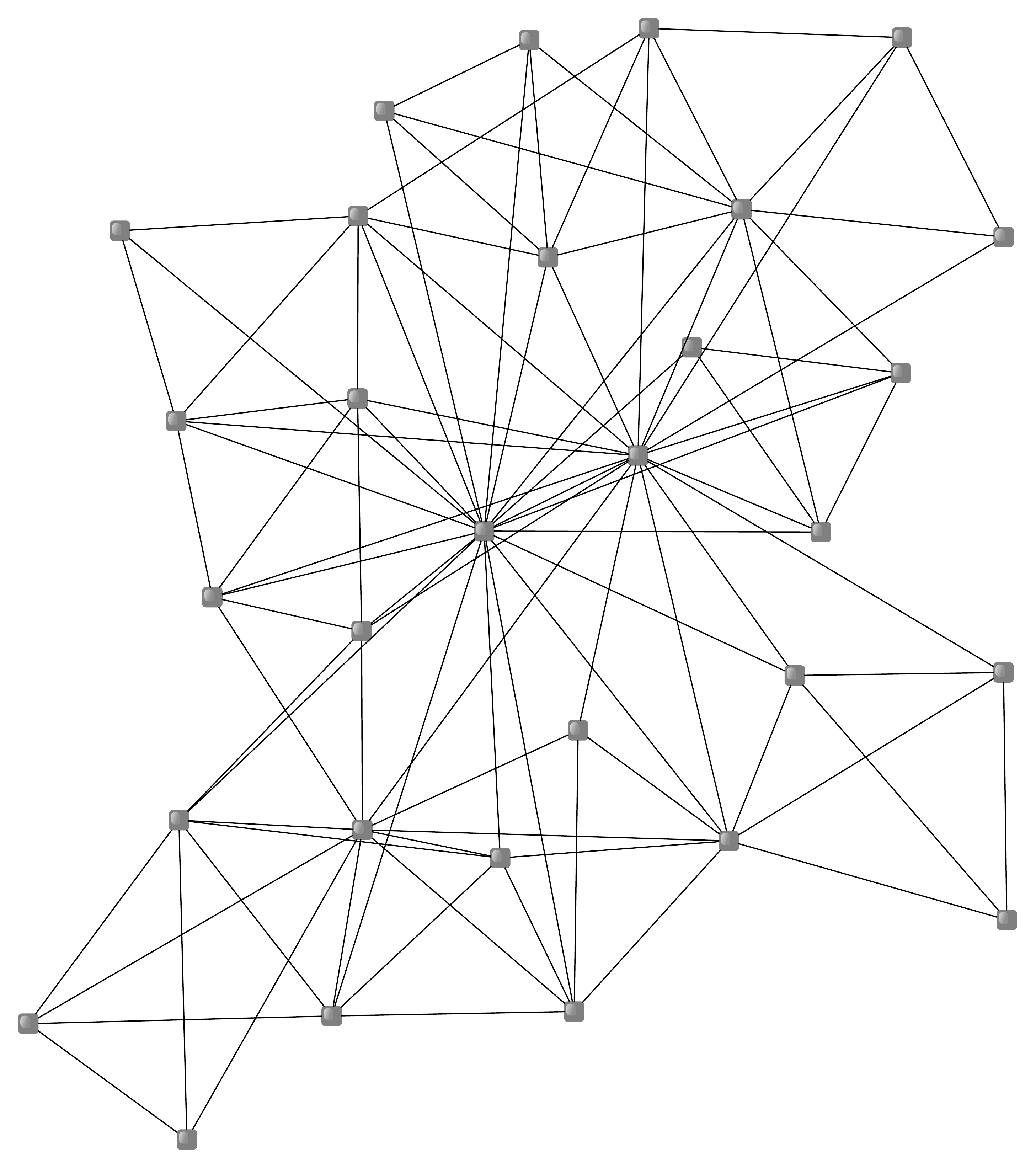}
}

\caption{Different drawings of the 5th graph given in the Graph Drawing 2017 contest produced by different variants of
(a)--(c)~the variant of our algorithm without restrictions on the aspect ratio, 
(d)--(f)~the variant of our algorithm forced to maintain the input aspect ratio,
(g)--(i)~the algorithm by Argyriou et al.~\cite{DBLP:journals/cj/ArgyriouBS13}, and
(j)--(l)~the algorithm by Huang et al.~\cite{DBLP:journals/vlc/HuangEHL13}.
The aesthetic criterion optimized by each variant is reported in the caption of its subfigure.}
\label{fig:graph5}
\end{figure}

\begin{figure}[htbp]
\centering
\subfloat[\label{fig:graph9-ourBest-cr}{Crossing Resolution}]{
\centering
\includegraphics[width=0.3\textwidth]{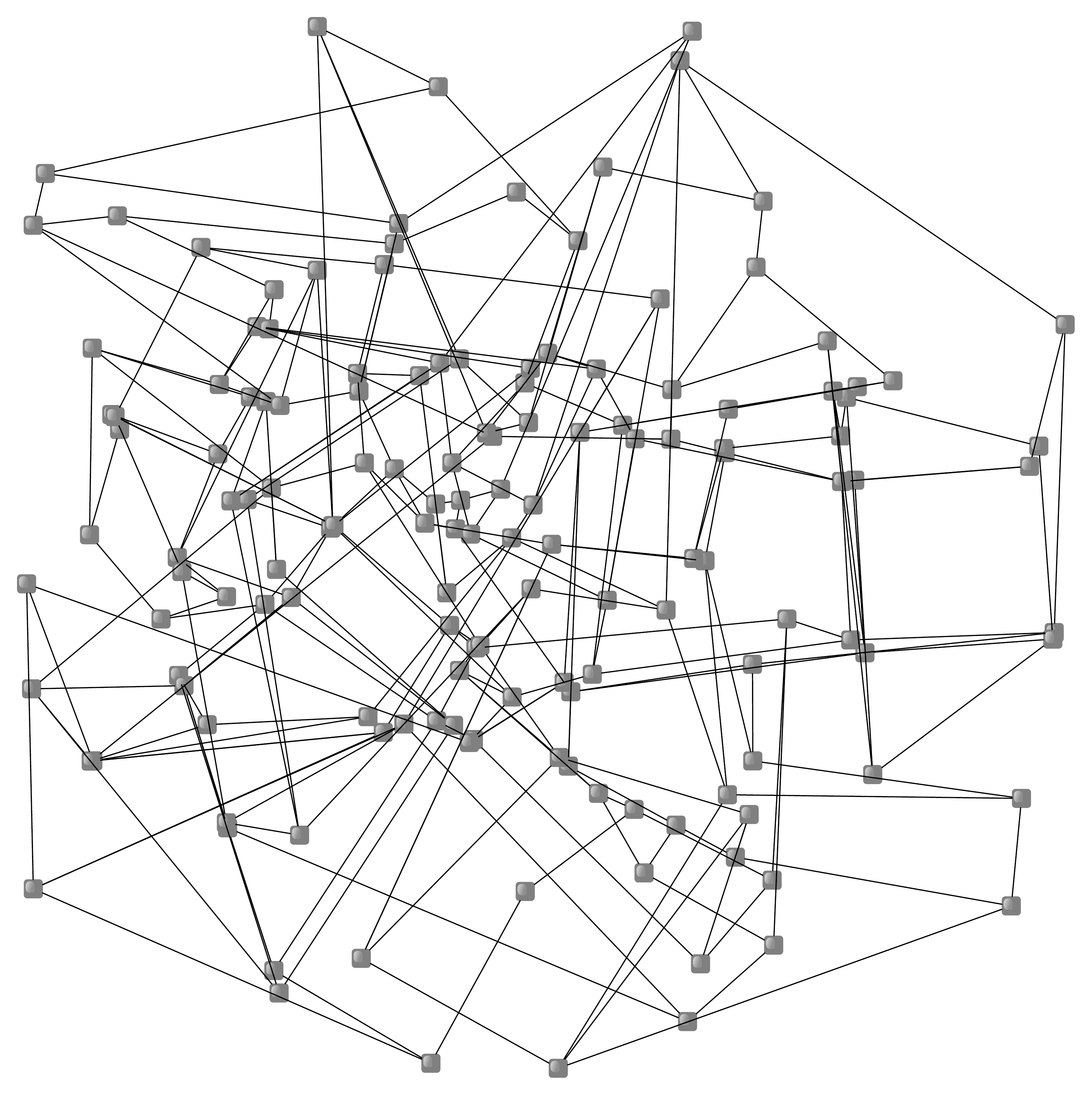}
}
\hfill
\subfloat[\label{fig:graph9-ourBest-ar}{Angular Resolution}]{
\centering
\includegraphics[width=0.3\textwidth]{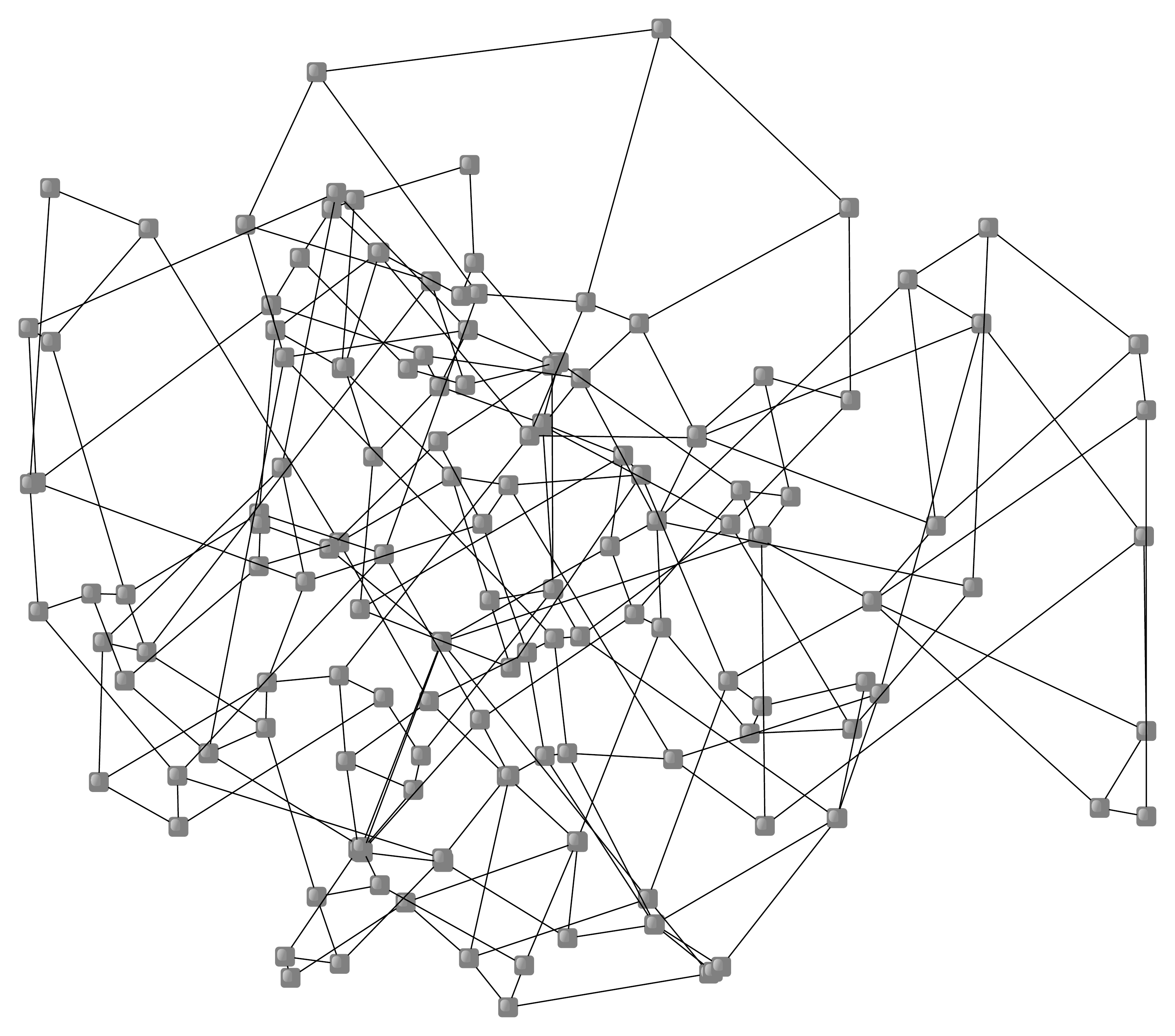}
}
\hfill
\subfloat[\label{fig:graph9-ourBest-tr}{Total Resolution}]{
\centering
\includegraphics[width=0.3\textwidth]{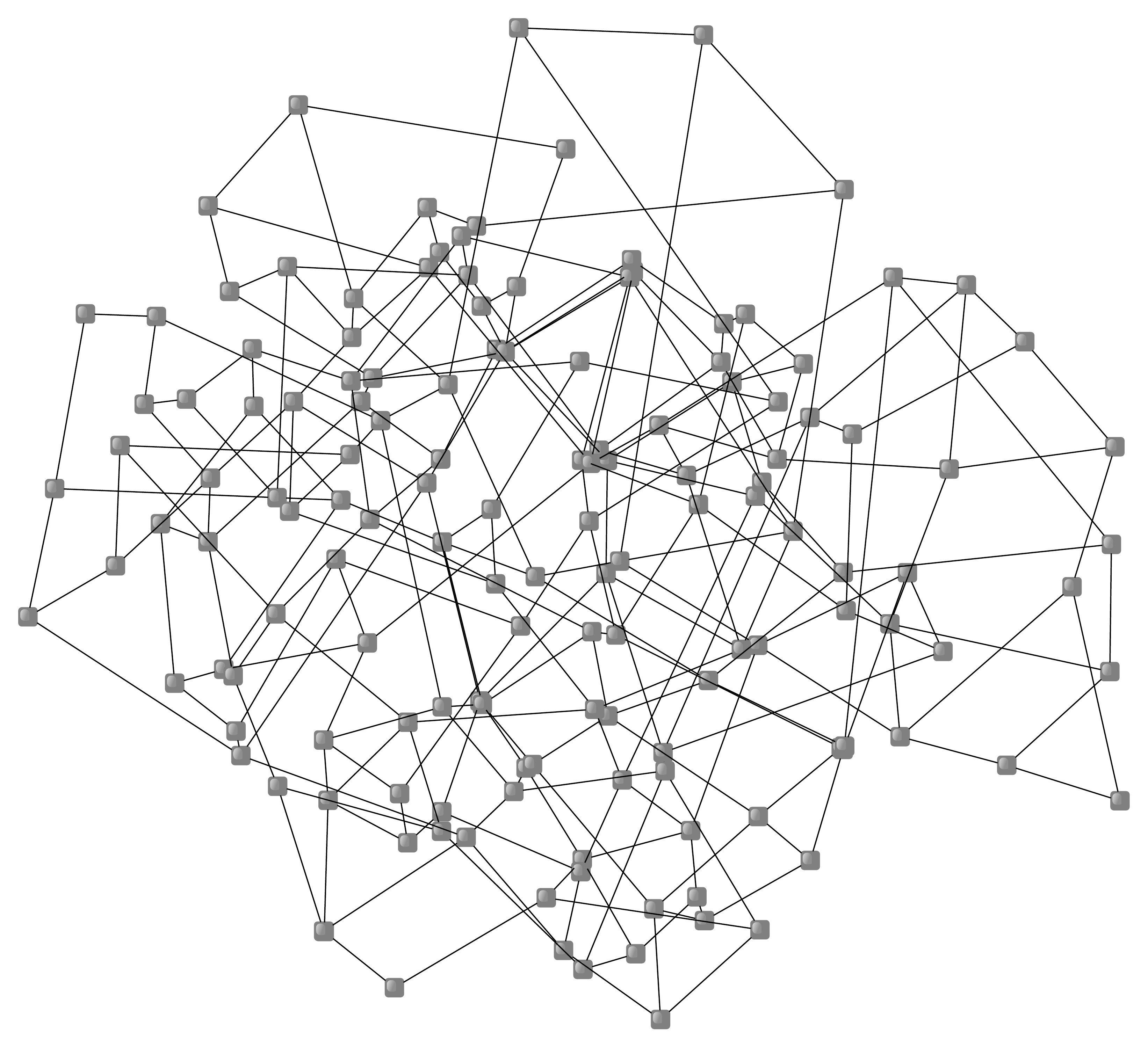}
}

\subfloat[\label{fig:graph9-ourRestricted-cr}{Crossing Resolution}]{
\centering
\includegraphics[width=0.3\textwidth]{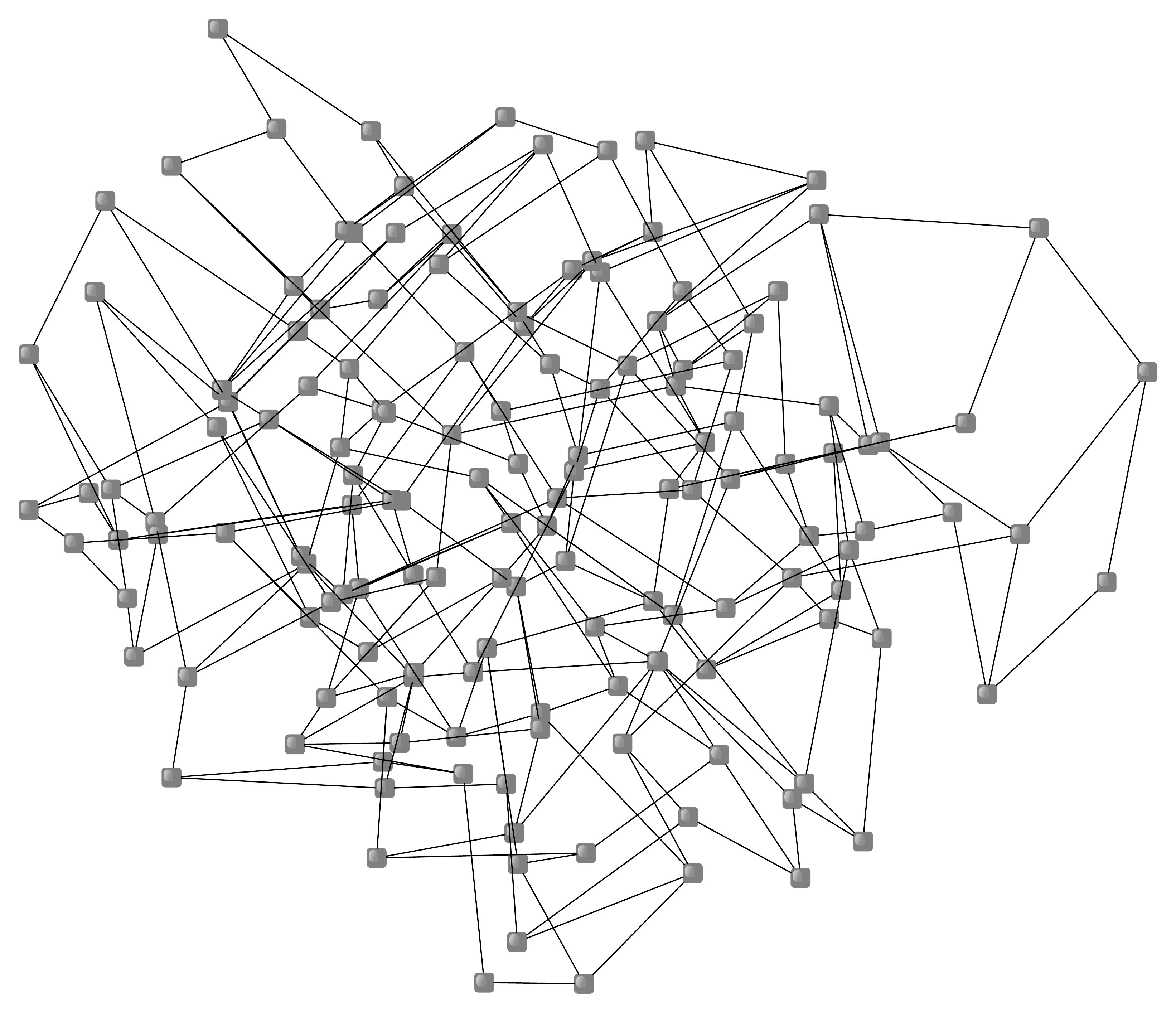}
}
\hfill
\subfloat[\label{fig:graph9-ourRestricted-ar}{Angular Resolution}]{
\centering
\includegraphics[width=0.3\textwidth]{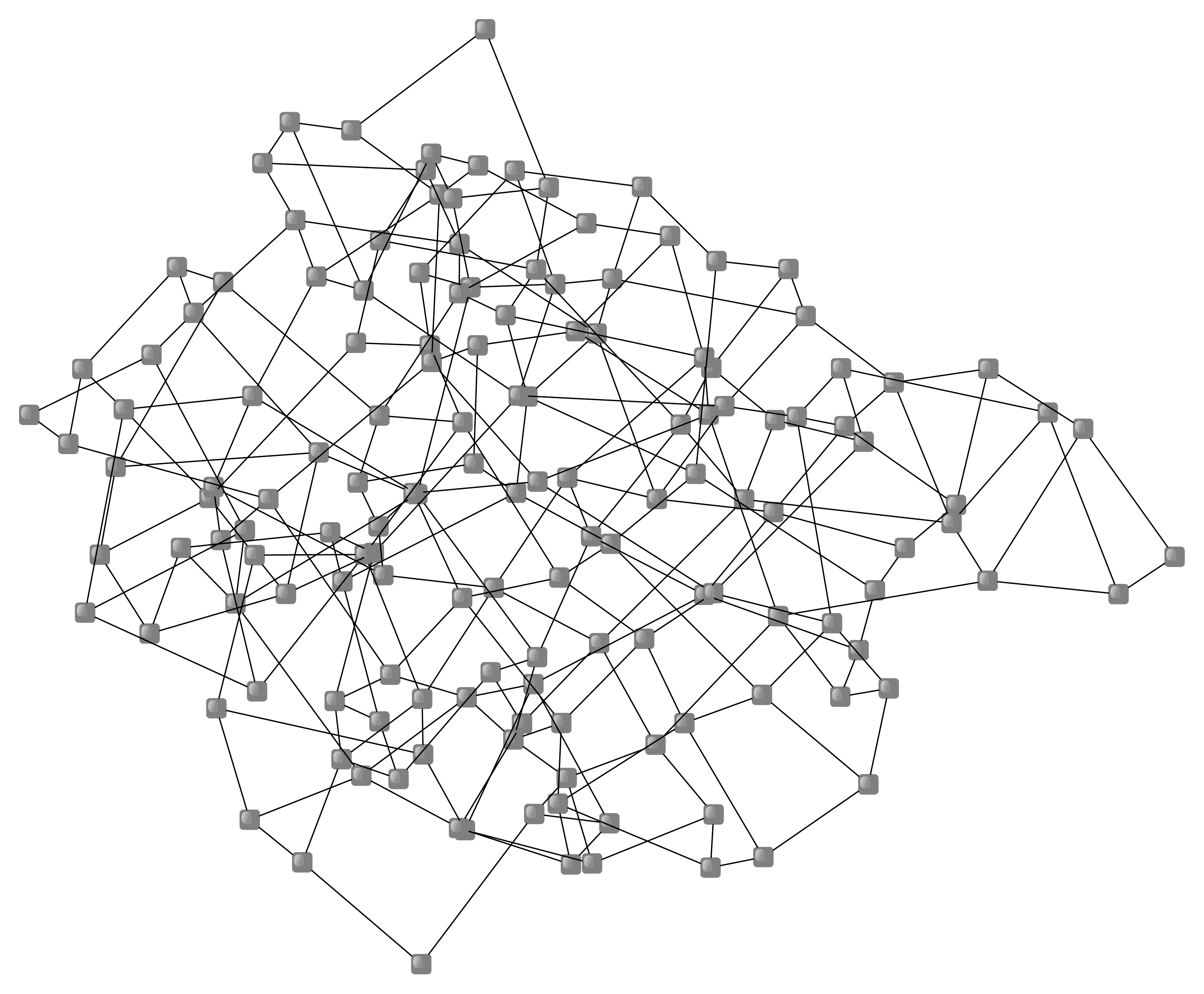}
}
\hfill
\subfloat[\label{fig:graph9-ourRestricted-tr}{Total Resolution}]{
\centering
\includegraphics[width=0.3\textwidth]{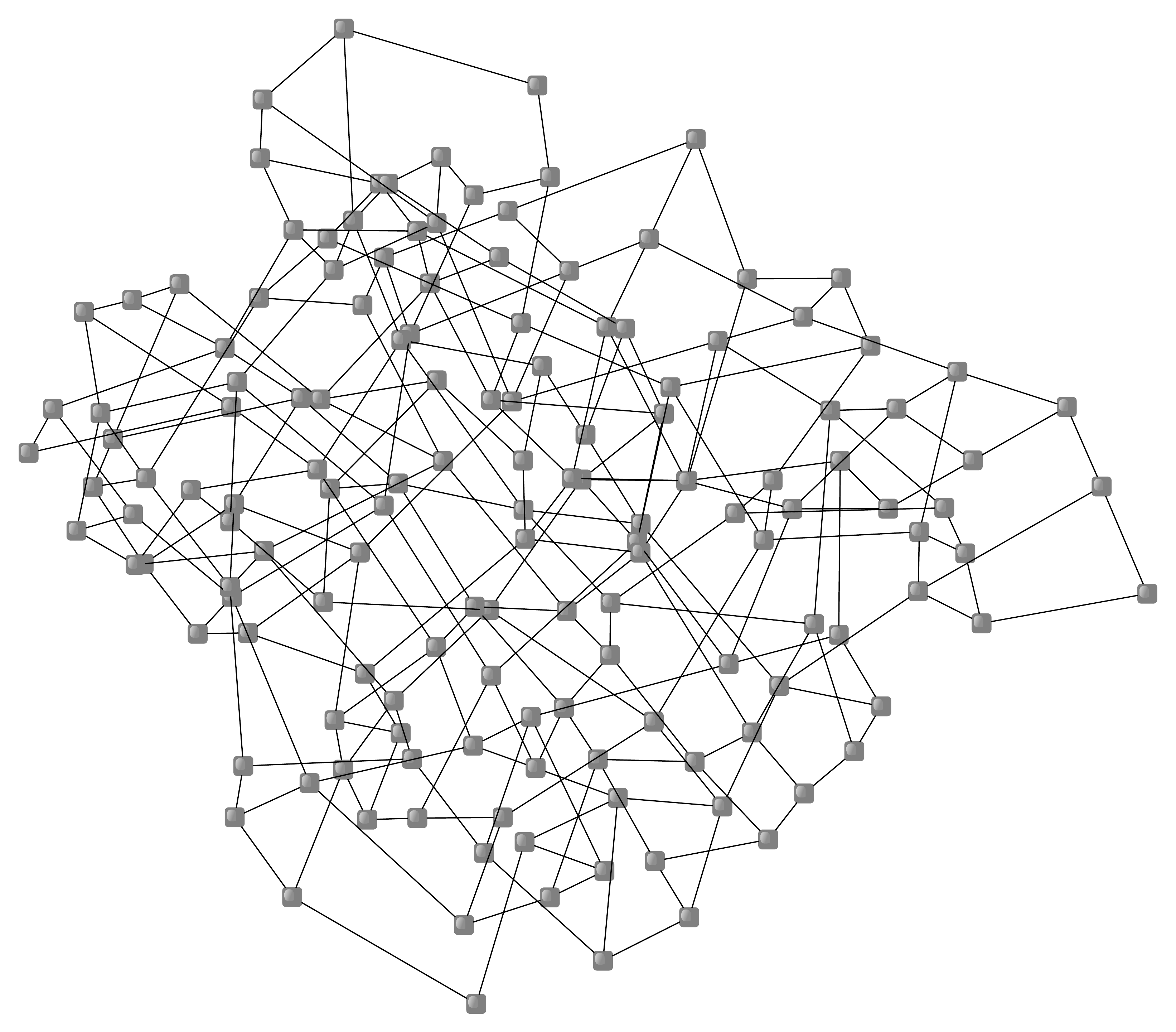}
}

\subfloat[\label{fig:graph9-argyriou-cr}{Crossing Resolution}]{
\centering
\includegraphics[width=0.3\textwidth]{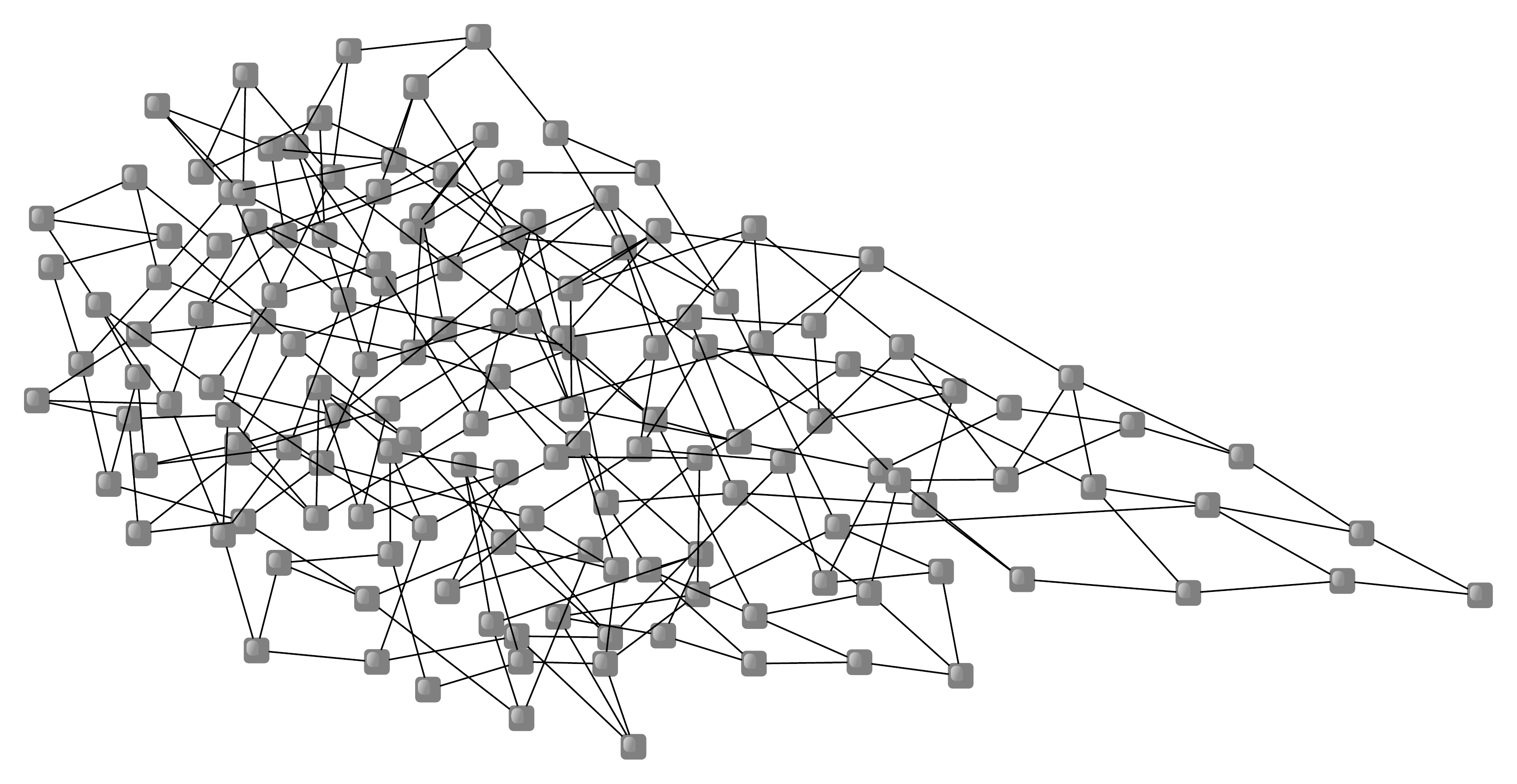}
}
\hfill
\subfloat[\label{fig:graph9-argyriou-ar}{Angular Resolution}]{
\centering
\includegraphics[width=0.3\textwidth]{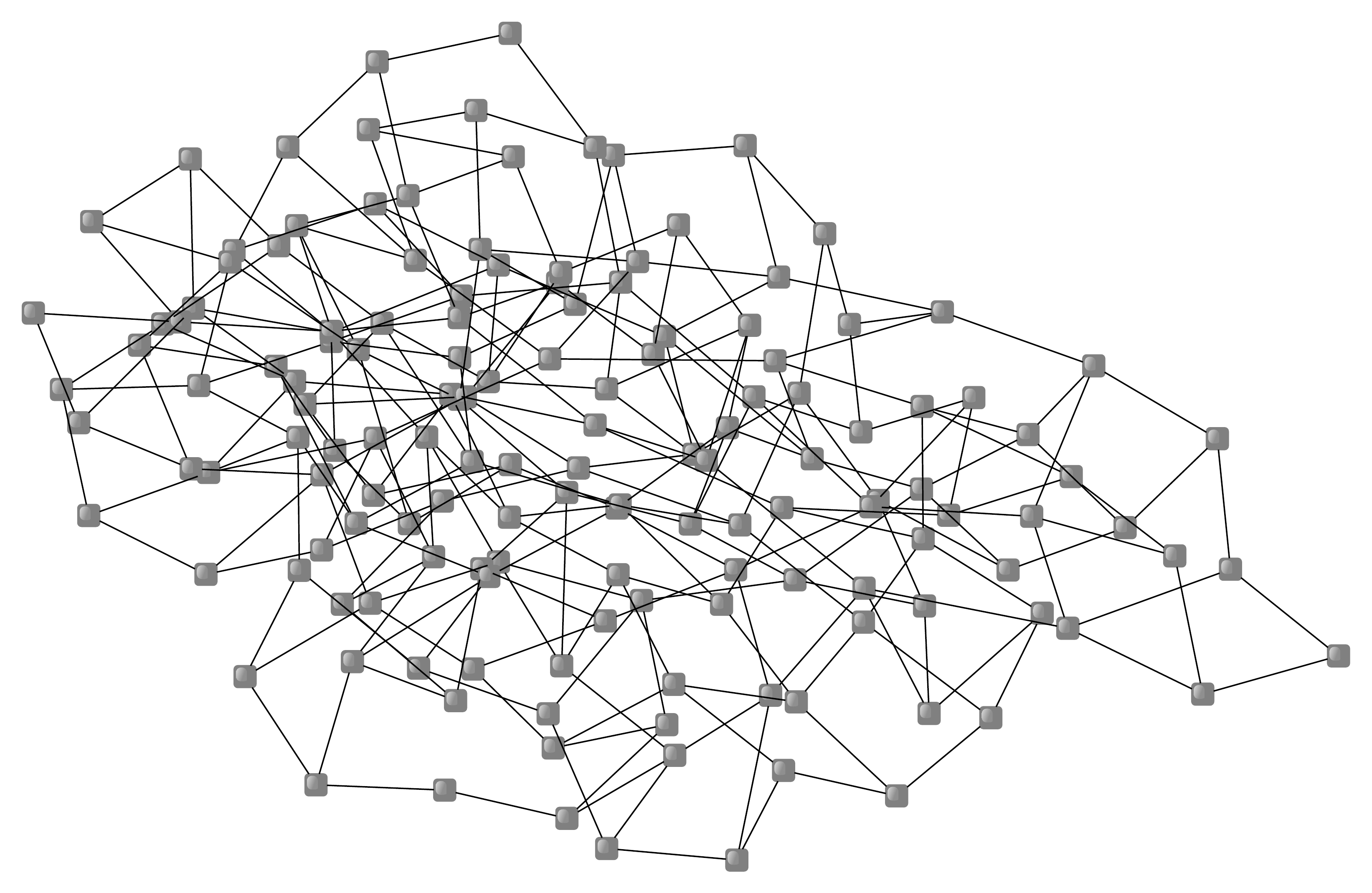}
}
\hfill
\subfloat[\label{fig:graph9-argyriou-tr}{Total Resolution}]{
\centering
\includegraphics[width=0.3\textwidth]{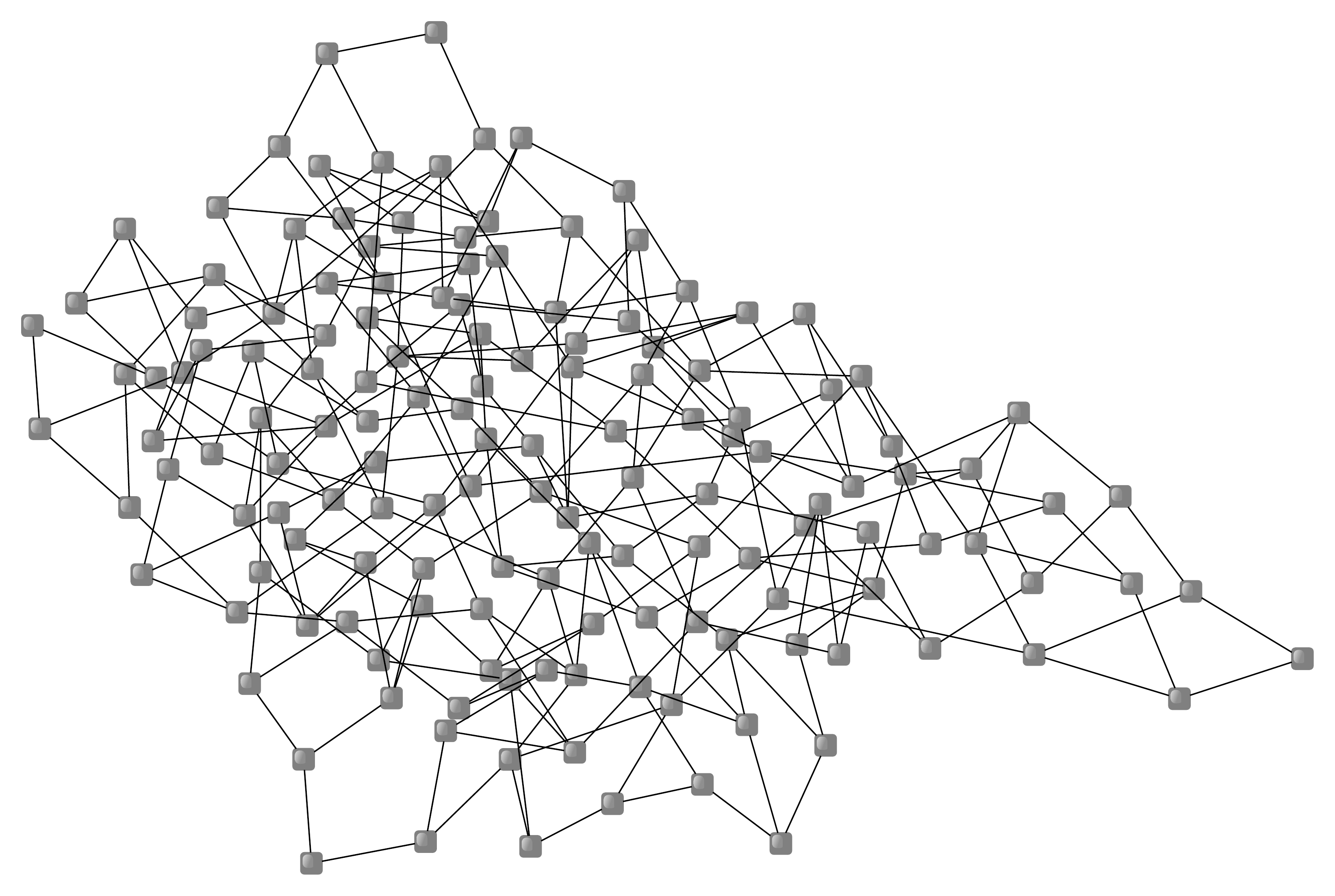}
}

\subfloat[\label{fig:graph9-huang-cr}{Crossing Resolution}]{
\centering
\includegraphics[width=0.3\textwidth]{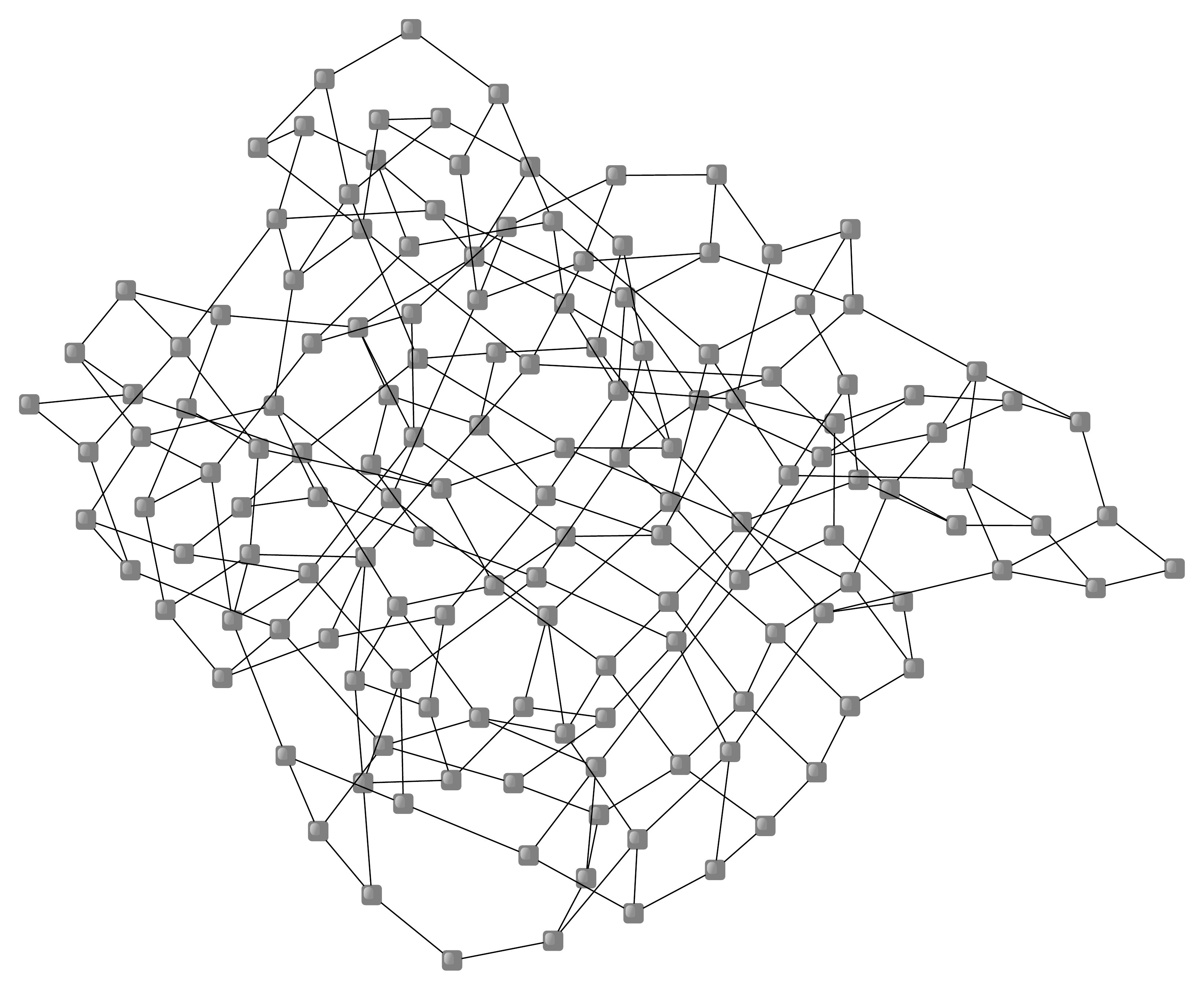}
}
\hfill
\subfloat[\label{fig:graph9-huang-ar}{Angular Resolution}]{
\centering
\includegraphics[width=0.3\textwidth]{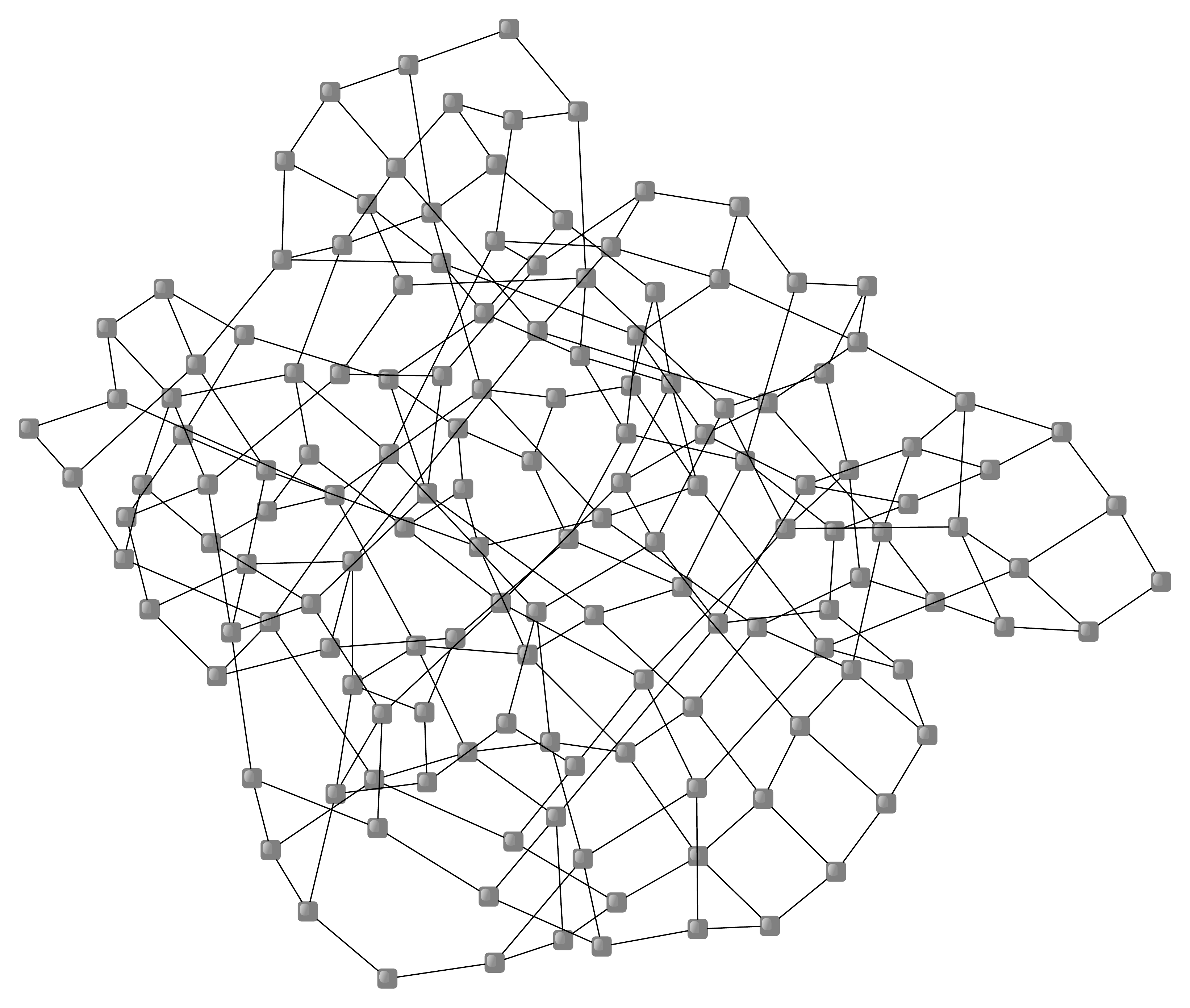}
}
\hfill
\subfloat[\label{fig:graph9-huang-tr}{Total Resolution}]{
\centering
\includegraphics[width=0.3\textwidth]{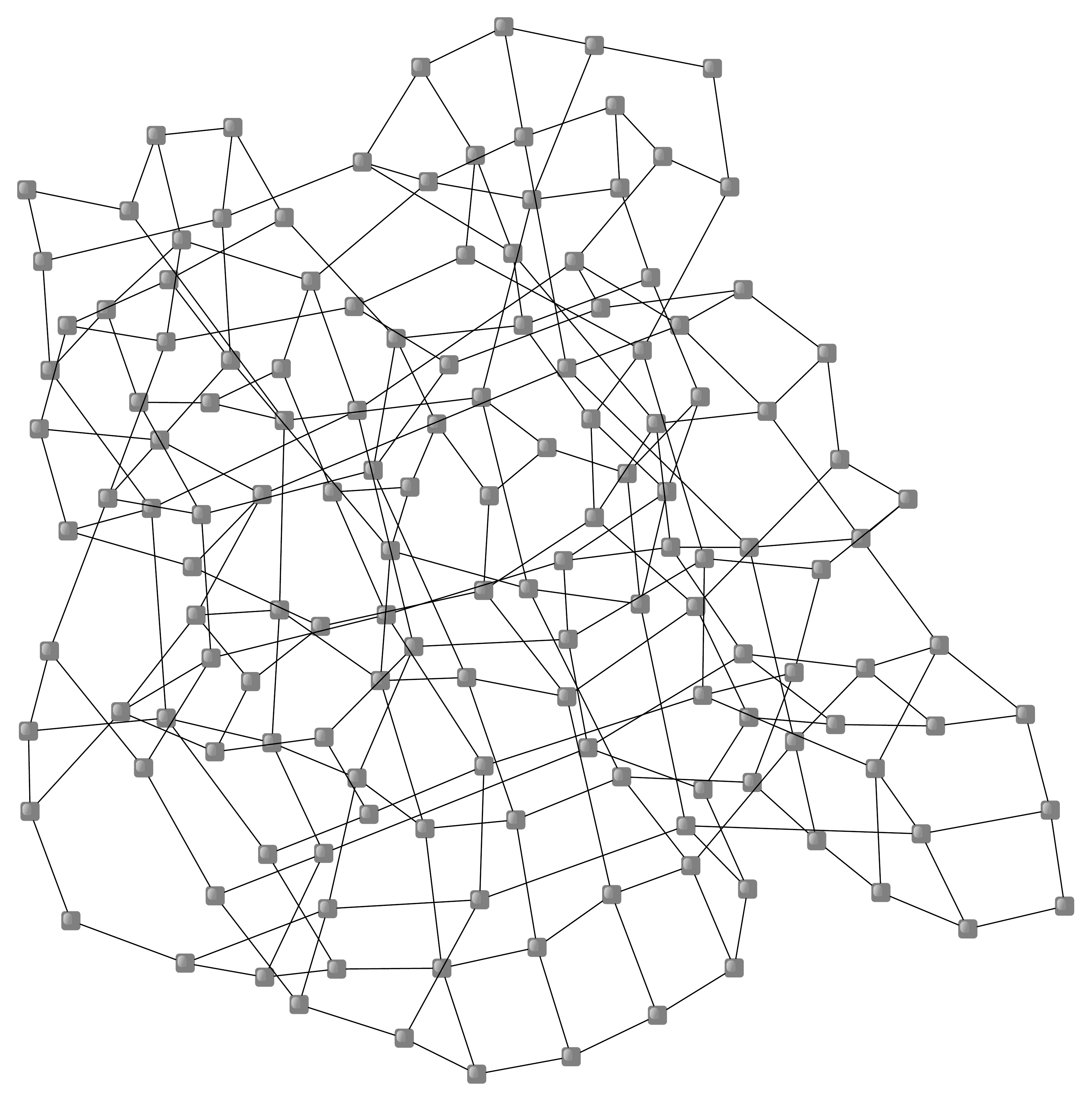}
}

\caption{Different drawings of the 9th graph given in the Graph Drawing 2017 contest produced by different variants of
(a)--(c)~the variant of our algorithm without restrictions on the aspect ratio, 
(d)--(f)~the variant of our algorithm forced to maintain the input aspect ratio,
(g)--(i)~the algorithm by Argyriou et al.~\cite{DBLP:journals/cj/ArgyriouBS13}, and
(j)--(l)~the algorithm by Huang et al.~\cite{DBLP:journals/vlc/HuangEHL13}.
Each variant was obtained by optimizing a different aesthetic criterion, which is named in the caption of each subfigure.}
\label{fig:graph9}
\end{figure}

}{}

\end{document}